\documentclass{article}

\usepackage{PRIMEarxiv}

\usepackage{biblatex}
\usepackage[utf8]{inputenc} % allow utf-8 input
\usepackage[T1]{fontenc}    % use 8-bit T1 fonts
\usepackage{hyperref}       % hyperlinks
\usepackage{url}            % simple URL typesetting
\usepackage{booktabs}       % professional-quality tables
\usepackage{amsfonts}       % blackboard math symbols
\usepackage{nicefrac}       % compact symbols for 1/2, etc.
\usepackage{microtype}      % microtypography
\usepackage{lipsum}
\usepackage{fancyhdr}       % header
\usepackage{graphicx}       % graphics
\graphicspath{{media/}}     % organize your images and other figures under media/ folder
\title{Efficient Design of RNA Sequences with Desired Properties, Structure, and Motifs Using A Grammar Variational Autoencoder
}

\author{
  Narges Zarnaghinaghsh \\
  Department of Electrical and Computer Engineering \\
  Texas A\&M University \\
  College Station, TX 77843, USA\\
  \texttt{nzarnaghi@tamu.edu} \\
   \And
  Byung-Jun Yoon \\
  Department of Electrical and Computer Engineering \\
  Texas A\&M University \\
  College Station, TX 77843, USA\\
  \texttt{bjyoon@tamu.edu} \\
}

\begin{document}
\maketitle

\begin{abstract}
Designing structurally stable RNA sequences with specific motifs and other desirable properties is an
important challenge in bioinformatics. The potential design space increases exponentially with the
length of the RNA to be engineered, which makes this a difficult combinatorial optimization problem.
In this paper, we propose an RNA grammar variational autoencoder (RGVAE) that can efficiently
generate novel RNA sequences with specific target properties. The proposed RGVAE builds on the
recently proposed grammar VAE, where we incorporate the stochastic context-free grammar (SCFG)
to design strutural RNAs with desired motifs and characteristics. Using the SCFG can ensure that
the generated RNA sequence can form a thermodynamically stable secondary structure. Given a
RNA sequence, the SCFT is used to find the parse tree, which is represented in a continuous low-
dimensional latent space by the RGVAE encoder. We can optimize the RNA in the latent space, where
the latent representation can be decoded by the RGVAE decoder to reconstruct the RNA sequence.
Based on a number of practical uses cases, we demonstrate that RGVAE can be used to efficiently
design structurally stable RNAs with specific target properties, which significantly outperform other
alternatives such as randomized design and regular VAEs that do not utilize the SCFG.
Code availability: the source code of RGVAE and the data used in this study are provided in https://github.com/nzarnaghinaghsh/RGVAE/tree/main, DOI 10.5281/zenodo.15569206.
%\lipsum[1]
\end{abstract}

% keywords can be removed

\section{Introduction}

The secondary structure of an RNA plays a crucial role in performing its biological functions\cite{bosecauses}. For this reason, the RNA design problem -- namely, to engineer RNAs that possess desired characteristics -- often involves predicting the structure of a designed RNA sequence or the inverse design of an RNA sequence that can fold into a specific structure, typically referred to as the inverse folding problem~\cite{eastman2018solving}. 
Designing RNA sequences by solving the inverse folding problem has important applications, for example, the design of RNAs for gene regulation that may be potentially useful in drug discovery ~\cite{esmaili2015erd}. Depending on the application, there may be other features, in addition to forming a specific secondary structure, that may be desired when designing RNA sequences. For example, this may include thermodynamic stability (e.g., measured in terms of the minimum free energy
structure, which could be predicted by using popular tools such as ViennaRNA~\cite{lorenz2011viennarna}), specific sequence motifs to include or avoid, or additional constraints based on the GC-content. Since the number of possible sequences increases exponentially with the size of the RNA to be designed, exhaustively searching the design space becomes quickly infeasible except for very short RNAs, and we need computationally efficient means for optimizing the RNA design to satisfy the constraints and to enhance the desired properties \cite{esmaili2015erd}.

% Deep learning (DL) algorithms employ different neural network (NN) architectures.  These algorithms have been widely used for numerous applications such as image, voice, object recognition and classification, image generation, and bioinformatics during the last decade. 

Recently, deep learning (DL) methods have been used for generating new chemical compounds with specific chemical properties~\cite{xu2019deep} and generative DL models are being actively explored for accelerating the discovery of novel drugs or materials~\cite{segler2018generating, yuan2017chemical,gomez2018automatic,blaschke2018application,karimi2020novo,kadurin2017drugan,kusner2017grammar}. For example, \cite{segler2018generating}~\cite{segler2018generating} and~\cite{yuan2017chemical}~\cite{yuan2017chemical} used RNN (Recurrent Neural Network) as a generative model for molecular structures. \cite{segler2018generating} generated new drug-like molecules using RNNs by sampling from a probability distribution learned from the training data. In~\cite{yuan2017chemical}, char-RNN (character-level RNN) was used to increase the structural diversity and the chemical novelty of the generated molecules. In \cite{iwano2022generative}, the authors developed a variational autoencoder to generate in silico aptamer. In \cite{runge2023towards}, the authors developed libLEARNA, which is a structure-based method for designing riboswitches with variable lengths and specific features. In \cite{franke2022probabilistic}, the authors propose a hierarchical latent distribution for improving the transformer models to modify the distribution of the data and the ambiguities. They provide the advantages of using their method for RNA folding. In \cite{bagal2021molgpt}, the authors present the generative model, MolGPT, for generating new valid molecules with specific properties.

Therefore, we propose a hierarchical latent distribution to enhance one of the most successful deep learning models, the Transformer, to accommodate ambiguities and data distributions. We show the benefits of our approach (1) on a synthetic task that captures the ability to learn a hidden data distribution, (2) with state-of-the-art results in RNA folding that reveal advantages on highly ambiguous data, and (3) demonstrating its generative capabilities on property-based molecule design by implicitly learning the underlying distributions and outperforming existing work.

% One of the DL models is convolutional neural networks (CNN) which are used in image classification problems~\cite{guo2017simple} or network intrusion detection~\cite{khan2019improved}. Recurrent neural networks (RNN) are another kind of neural networks which are used in different applications that use time-series or sequence data such as text classification~\cite{liu2016recurrent}. Long short-term memories (LSTM) networks are a set of RNN networks that processes the entire sequence of data. LSTM networks are used in many applications such as large-scale acoustic modeling~\cite{sak2014long}. Another type of RNNs are gated recurrent units (GRUs) which are used in many applications including speech recognition~\cite{ravanelli2017improving}, forecasting of COVID-19~\cite{arunkumar2021forecasting}, or predicting prices of stock market~\cite{rahman2019predicting}.

% Generating novel structures by using generative models is another important application of DL.~\cite{segler2018generating} and~\cite{yuan2017chemical} used RNN as a generative model for molecular structures.~\cite{segler2018generating} uses a model that generates new drug-like molecules by sampling from a probability distribution learned from the training data. In~\cite{yuan2017chemical}, the structural diversity and the chemical novelty of the generated molecules using the generative model increases.

%Autoencoder (AE) is a type of NN that encodes the input data to a latent space and then decodes from the latent space to regenerate the input data

\cite{gomez2018automatic}~\cite{gomez2018automatic} proposed the use of a variational autoencoder (VAE) for generative molecular design. The VAE's encoder maps molecules to a continuous and low-dimensional latent space, thereby addressing the computational challenges of optimizing molecules in the original molecular space, which is discrete and high-dimensional. This makes the optimization problem more amenable and can improve the efficiency of predicting novel molecules with enhanced properties. Once new molecules are sampled in the latent space, they can be decoded back to the original molecular space by the VAE's decoder. Similarly,
\cite{blaschke2018application}~\cite{blaschke2018application} used a generative autoencoder for \textit{de novo} molecular design. Molecular structures are encoded into a latent space, which is investigated to produce novel structures with desired characteristics, and then decoded back to the original molecular space. \cite{karimi2020novo}~\cite{karimi2020novo} used another type of deep generative model -- namely, a semisupervised gcWGAN (guided, conditional, Wasserstein Generative Adversarial Networks) -- to find the sequence-structure relationships. The resulting generative model was used to design protein sequences for novel structural folds. \cite{kadurin2017drugan}~\cite{kadurin2017drugan} used a generative adversarial autoencoder (AAE) for molecular feature extraction applications. They showed that their proposed AAE model notably improves the design of new molecules with targeted anticancer properties.

Another interesting variant of the VAE is the grammar VAE (GVAE)~\cite{kusner2017grammar}, which combines a transformational grammar with a VAE. By utilizing a grammar to parse and decode sequences, the GVAE can mask out invalid generated sequences that do not follow the grammar rules, thereby improving the validity of the generated sequences.
\cite{kusner2017grammar} demonstrated how a GVAE can be used for efficiently designing valid molecules with specific optimized features. Similarly, \cite{kraev2018grammars}~\cite{kraev2018grammars} combined grammars with reinforcement learning (RL) for designing novel molecules that possess specific desired characteristics. \cite{sumi2024deep} suggests RfamGen which is a generative model for designing RNA family sequences.  In addition, this paper provides results showing the high capability of GVAE and RfamGen for generating RNA sequences.

To date, various RNA sequence generation and design methods have been proposed in the literature~\cite{hammer2019evolving}. For example, \cite{esmaili2015erd}~\cite{esmaili2015erd} proposed an algorithm for RNA design with various energy and sequence  constraints. The energy constraint imposes a specific range on the free energy that the designed RNAs can take whereas the sequence constraints enforce certain sequence positions to have predefined nucleotides. \cite{zhou2013flexible}~\cite{zhou2013flexible} developed a method for RNA design with specific position and motifs constraints using a context-free grammar (CFG). More recently, \cite{eastman2018solving}~\cite{eastman2018solving} proposed a method for computational RNA design using RL. 
In principle, the RNA design problem can also take advantage of generative DL models for effective and computationally efficient RNA design with various constraints and target properties.

In this paper, motivated by the previous successful applications of DL algorithms for generating molecular structures and sequences, we propose a novel deep-learning-based approach for efficient RNA design. Specifically, we build on the grammar VAE~\cite{kusner2017grammar} to propose an RNA grammar variational autoencoder (RGVAE) that combines the descriptive power of stochastic context-free grammars (SCFGs) for modeling RNA sequences with underlying secondary structures with the generative capability of VAEs for molecular design. Although \cite{sumi2024deep} discusses GVAE for RNA sequences, we use SCFGs rather than CFG, and the grammar, applications, design features, and optimization are completely different with our proposed method. Based on the proposed RGVAE, we demonstrate that the model enables computationally efficient optimization of RNA design under multiple constraints. In this study, we consider several widely considered constraints in RNA design and properties -- including, minimum free energy (MFE),  GC content, target secondary structure, range of RNA size, base positional constraints, and mandatory or forbidden sequence motifs. Under diverse design scenarios that impose various subsets of the aforementioned constraints, we show that the proposed RGVAE is highly flexible and is capable of generating novel RNA sequences with the desired characteristics.

% with VAE for the design and generation of RNA sequences with particular constraints. These constraints are MFE minimization, a target GC content, a target secondary structure, a range for RNA length, base positional constraints, and mandatory and forbidden motifs. We investigate different combinations of these constraints. MFE minimization for RNA sequence generation has not been discussed before in the literature. In addition, the investigation of the combination of all of these constraints together is novel. We present a variety of scenarios for different combinations of these constraints. Furthermore, our proposed model is trained for tRNA sequences. Hence, the generated RNA sequences have specific secondary structures.Finally, in order to show the applicability of our suggested method, we generate RNA sequences for realistic applications.

\section{Methods}
In this paper, we use RGVAE for the RNA sequence generation and design. This model combines SCFG with VAE to ensure the validity of the generated sequences. We first review a number of concepts and existing models in the literature, on which the proposed framework is built. These include the variational autoencoder (VAE), context-free grammar (CFG), stochastic context-free grammar (SCFG), and grammar VAE (GVAE). Next, we propose the RNA Grammar Variational Autoencoder (RGVAE) and describe how it can be used for designing novel RNA sequences with desired properties. Using the RGVAE and Bayesian optimization, we optimize the  constraints to improve the desired features. The summation of the scores of different constraints for Bayesian optimization is implemented. The data for the optimization is generated using the RGVAE generative model. 

\subsection{Variational autoencoder (VAE)}
An autoencoder includes two parts: an encoder deep  network to transform the input sequences into a vector, and a deep decoder network to transform the vectors back into the strings \cite{cemgil2020autoencoding}. The goal of the training of an autoencoder is to minimize the error to regenerate the primary strings~\cite{gomez2018automatic}.  The vector-encoded string is called the \textit{latent representation} of that string \cite{pinheiro2021variational}.

To avoid invalid decoding for unconstrained optimization in the latent space, decoded strings must have a chemical nature similar to the training data. Otherwise, the latent space may have huge "dead areas". It is important to notice that these "dead areas" cause invalid decoded sequences \cite{gomez2018automatic}. Therefore, in order to avoid this problem, a \textit{variational autoencoder} (VAE) was developed. VAE adds the combination of stochasticity with a penalty term to the encoder in an autoencoder to avoid invalid decoding. That is because adding noise to the encoded sequences helps to train the decoder to decode a wider range of latent points.

\subsection{Context-free grammar (CFG)}
A context free grammar (CFG) in \cite{dowell2004evaluation} and ~\cite{durbin1998biological} is defined as \textit{G=(V,T,P,S)} where:
\begin{enumerate}
\item[(1)] \textit{V} is a finite set of non-terminal symbols;

\item[(2)] \textit{T} is a finite set of terminal symbols (For RNA sequences, these terminal symbols are \textit{\{a,c,g,u\}});

\item[(3)] \textit{P} is a finite set of production rules; and
\item[(4)] \textit{S} is the start variable which is a non-terminal symbol which is named as the \textit{start symbol};
\end{enumerate}
Starting from the start symbol, production rules are used to explain how the grammar produces an observed symbol string step by step. The production rules are described as $B \rightarrow \alpha$ where $B \in V$ is a left-hand side non-terminal symbol and $\alpha \in \left(V \cup T\right)^*$ is a sequence of terminal and/or non-terminal symbols. As an example, $S \rightarrow ASU$ is a CFG production rule for RNA sequences. This rule produces the terminal symbols `$A$' and `$U$'. The generated sequences of this grammar consist only of symbols $A$ and $B$.  

We use the production rules to transform an RNA sequence into a sequence of terminal symbols. Starting from the start symbol $S$, at each step, we choose a non-terminal symbol and transform it to a new series of terminal symbols and/or non-terminal symbols using the production rules. We continue the steps using the production rules to reach a sequence of just the terminal symbols. 

For instance, consider the grammar $G_{example}=(\{S\},\{a,b\},\{S \rightarrow aSa, S \rightarrow bSb, S \rightarrow \epsilon\},S)$ which only produces symbols containing $a$ and $b$. An example derivation of this grammar is $S \rightarrow bSb \rightarrow baSab \rightarrow babSbab \rightarrow babbab$ which generates the string $babbab$. 

By using the production rules recursively to non-terminal symbols, a group of possible trees are defined. In these trees, the left-hand-side symbols are the parents of the right-hand side child. At each step, a parent non-terminal symbol generates a terminal and/or non-terminal child symbol until all leaves of the tree become the terminal symbols. Then, the language of the grammar $G$ is defined as the group of all possible arrangements of the terminal symbols which could be generated in a tree. A \textit{parse tree} is a tree that has a root at $S$ and an arrangement of terminal symbols as its leaves given a sequence of terminal symbols in the language.

\begin{figure}
\begin{center}
\includegraphics[width=1.5in]{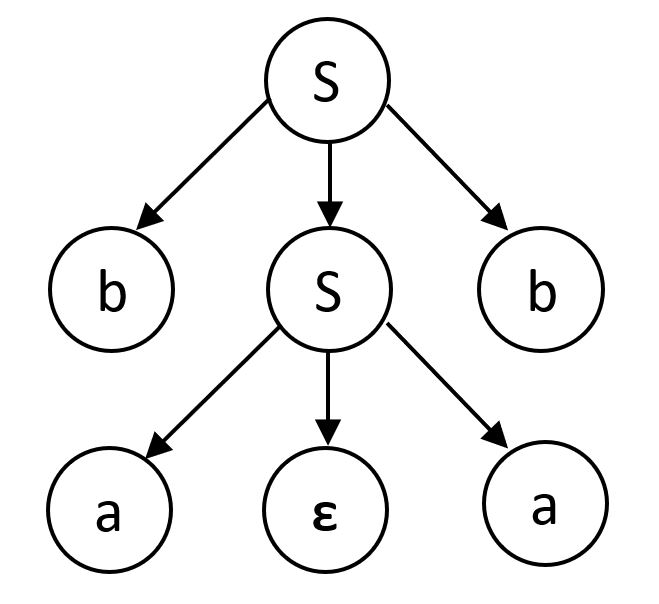}
\end{center}
\caption{An illustrative example of a parse tree for the sequence ``baab'' based on the grammar $G_{example}$.\label{F7}}
\end{figure}

An example of a parse tree for the grammar $G_{example}$ is depicted in Figure \ref{F7}.

\subsection{Stochastic context-free grammar (SCFG)}

In a stochastic context-free grammar (SCFG) G, a probability is assigned to each CFG production rule, and defining a probability distribution for parse trees is possible. The summation of the probabilities of the set of production rules from any non-terminal symbol must be one. Cocke-Younger-Kasami (CYK) dynamic programming
algorithm and Inside-Outside algorithm are proposed in the literature for finding the probabilities for SCFG~\cite{dowell2004evaluation}~\cite{eisner2016inside}. In this paper, we use the inside-outside algorithm for finding the probabilities for SCFG.

\subsection{Grammar variational autoencoder (GVAE)}

One of the problems with VAE is that the decoded sequences from the latent space may not be always valid. The Grammar Variational Autoencoder (GVAE) which is proposed in~\cite{kusner2017grammar} tries to add knowledge about generating valid sequences to VAE by combining a CFG with VAE. Using a CFG, any valid sequence could be parsed into a series of production rules. The GVAE tries to learn a VAE to generate a series of production rules. The advantage is that the valid sequences are generated using valid series of production rules.

\subsection{RNA grammar variational autoencoder (RGVAE)}

In this section, we provide the framework of RGVAE which we used for generating and optimizing the features of RNA sequences and explain how the grammar improves the performance of a variational autoencoder for generating valid RNA sequences. We closely follow the grammar variational autoencoder used in~\cite{kusner2017grammar}, but instead of a CFG, we use SCFG in our model. Furthermore, we upgraded the version of the code from Python 2.6 to Python 3, and we applied the model to different applications and constraints. As it is explained in ~\cite{kusner2017grammar},  SCFG could be useful for generated valid sequences.

In this paper, we use the following grammar which was used in~\cite{dowell2004evaluation}:

$G: S \rightarrow LS|L
$

$
     L \rightarrow aF\hat{a}|a
$

$
     F \rightarrow aF\hat{a}|LS
$

where $G$ is the grammar, $S$ is the start symbol, $L$ and $F$ are non-terminal symbols, $a$ shows any single terminal, and $\hat{a}$ shows the basepair of $a$. In our proposed framework, before using the encoder, we find the probabilities of the production rules by using the inside-outside algorithm. We apply this algorithm on the same data used for training the model. Then, we encode the RNA sequences into a latent space. Finally, we decode the points in the latent space to generate RNA sequences. Now, we describe the encoder, the decoder, and the training procedure of the proposed model in detail.

\subsubsection{RGVAE Encoder}

\begin{figure*}
%\setlength{\fBoxsep}{0pt}%
%\setlength{\fBoxrule}{0pt}%
%\begin{center}
\includegraphics[scale=0.35]{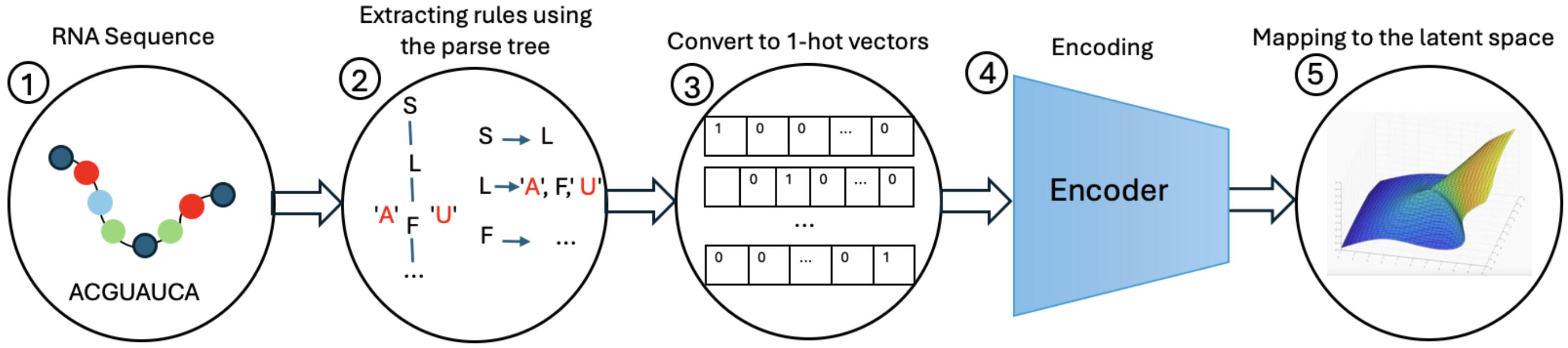}
%\end{center}
\caption{Illustration of the RGVAE encoding process. The overall architecture follows that of the grammar VAE~\cite{kusner2017grammar}, where we use a stochastic context-free grammar (SCFG) for parsing the RNA in order to faithfully represent the underlying secondary structure. Box 1 shows an RNA sequence. BOX 2 shows that given the RNA sequence, the SCFG is used to find the parse-tree, based on which the production rules are extracted and then converted into 1-hot vectors in Box 3. In Box 4, a CNN encoder is used to convert the 1-hot vectors to the latent space which is shown in Box 5. \label{F1}}
\end{figure*}

The structure of the encoding process is shown in Figure \ref{F1}. This structure is similar to the encoding process in~\cite{kusner2017grammar}, but instead of CFG, SCFG is used by finding the probabilities of SCFG, and a different grammar for RNA sequences is used instead of the grammar used in~\cite{kusner2017grammar} for molecules. Box 1 in this Figure shows an RNA sequence as the input. We used the production rules of the grammar $G$ to find a parse tree and the extract the rules of the RNA sequence in Box 2 which shows how the grammar produced an RNA sequence. In Box 3, these production rules are transformed into 1-hot vectors. Each of these vectors corresponds to an RNA grammar production rule. Then, in Box 4, a deep convolutional neural network (CNN) which is the same as the encoder in~\cite{kusner2017grammar} is used to transform the 1-hot indicator vectors to a continuous latent space. Let \textbf{z} be the latent vector in the latent space which is transformed from the 1-hot vector.
%previous: encoder.png
%\begin{widetext} 
%\end{widetext}

\subsubsection{RGVAE Decoder}

\begin{figure*}
\begin{center}
\includegraphics[scale=0.33]{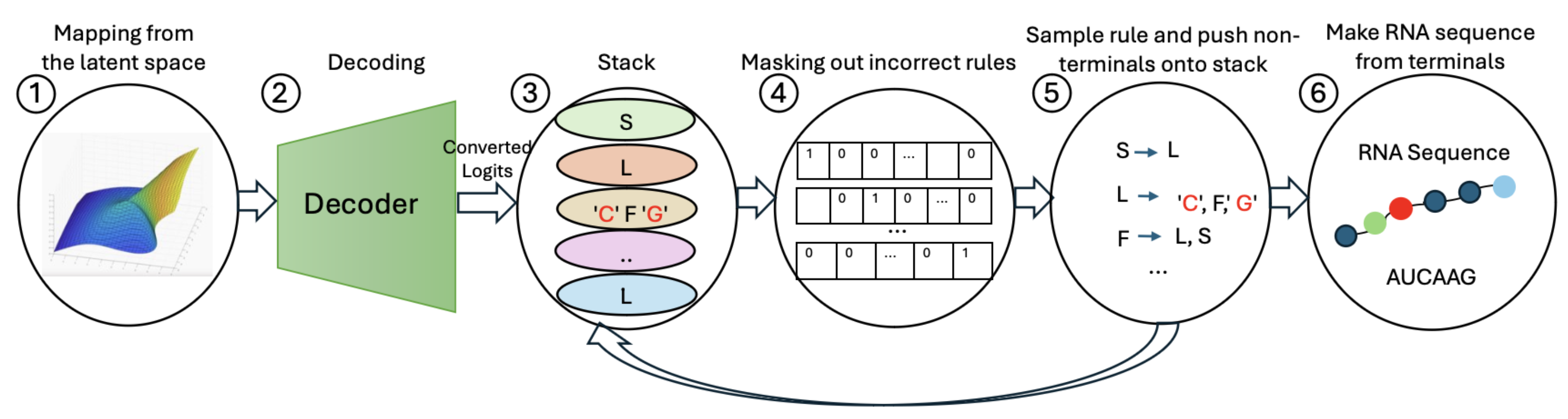}
\end{center}
\caption{Illustration of the RGVAE decoding process. The overall architecture of the decoder follows that of the grammar VAE~\cite{kusner2017grammar}. The latent space representation is mapped to the RNA sequence by applying the sampled production rules in the SCFG. The latent space is shown in Box 1. Box 2 shows an RNN decoder which generates logits. A stack is used in Box 3 to ensure the validity of the generated production rules. Then, we mask out incorrect rules in Box 4. In Box 5, we sample the uncovered rule. Boz 6 shows the generated RNA sequence. \label{F2}}
\end{figure*}

The structure of the decoding process is shown in Figure \ref{F2}. This is similar to the decoding process in~\cite{kusner2017grammar}, but the grammar and the application is different. It is important to mention that during the decoding process, we have a constraint on the decoder to choose only the `valid' production rules which ensures the generation of `valid' RNA sequences. We define the valid production rules as the set of the production rules of the selected SCFG for the training. 

As it is shown in Figure \ref{F2}, in Boxes 1 and 2, the continuous latent vector \textbf{z} is fed into an RNN that generates unnormalized log probability vectors (or `logits'). Each one of these logit vectors corresponds to a production rule which is similar to the 1-hot indicator vectors in the encoder part. In the following steps, these logit vectors are used to choose only the valid production rules.

In Figure \ref{F2}, in Box 3, we use a last-in fist-out stack similar to ~\cite{kusner2017grammar} in order to ensure the validity of the generated production rules. The beginning symbol for every valid parse should be the start symbol $S$. Hence, $S$ is put last in the stack. Then, we select any non-terminal symbol from the top of the stack (last symbol)  to cover the incorrect elements of the logit vector. A fixed binary mask vector $\textbf{m}_{\alpha}\in\left[0,1\right]^L$ is defined for any non-terminal $\alpha$. If a production rule has  $\alpha$ on their left side, all indices of the mask vector are `1'.

Because the start symbol $S$ is placed last on top of the stack, we first pick the start symbol $S$. For the symbol $S$, just the first production rule begins with $S$. Therefore, we put `1' for the first dimension and `0' for other dimensions as shown in Figure \ref{F2}, Box 4. Then, we use the masked distribution in Equation \ref{E1} to sample the other uncovered rules by utilizing their values in the logit vector at any timestep $t$ in Box 5:
\begin{equation}
    p\left(\textbf{X}_t =l|\alpha,\textbf{z}\right)=\frac{m_{\alpha,l}\exp\left(f_{tl}\right)}{\Sigma_{j=1}^Lm_{\alpha,l}\exp\left(f_{tj}\right)}
    \label{E1}
\end{equation}
where $f_{tl}$ is the $(t,l)$-component of the logit matrix \textbf{F}. 

The $S\rightarrow L$ will be used as the first production rule because just the first rule is unmasked. Then, the next rule begins with $L$. Hence, it is pushed on top of the stack as shown in Figure \ref{F2} in Box 3. One more time, we sample this non-terminal symbol and utilize it for covering every invalid rules in the logit vector. After that, a valid rule from this logit vector is sampled which is $L \rightarrow aF\hat{a}|a$. Again, the non-terminal symbols which are on the right side of this rule will be pushed on top of the stack. Then, we add the non-terminal symbols for those rules from right side to left side on the stack so that the leftmost non-terminal symbol will be the topmost symbol of the stack. Again, we pick the last rule from the top of the stack and cover the components in the logit vector. This procedure is continued until there is no symbol in the stack or the maximum number of logit vectors is reached. Then, we make the RNA sequences in Box 6 from the terminals of Box 5.

%previous image: decoder.eps

\subsection{Training of the RGVAE}

For training, let us assume that the maximum number of the production rules (timesteps) is denoted as $T_{\textrm{max}}$ . As we explained before, to encode an RNA sequence, the sequence is converted into 1-hot vectors $\textbf{X} \in \{0,1\}^{T_{\textrm{max}} \times L}$ which describes a series of $T_{\textrm{max}}$ mask vectors as well. For training, using the left side of the rule represented in the one-hot vector $\textbf{x}_t$ at timestep $t=1,..., T_{\textrm{max}}$, each mask is chosen. Using these masks, the mapping of the decoder is computed in Equation \ref{E2}:

\begin{equation}
    p(\textbf{X}|\textbf{z})= \prod_{t=1}^{T(\textbf{x})}p(\textbf{x}_t|\textbf{z}) 
    \label{E2}
\end{equation}

$T\left(\textbf{x}\right)$ is the total number of production rules (timesteps) used to generate an output sequence for $\textbf{x}$. $p(\textbf{x}_t|\textbf{z})$ in equation \ref{E2} is obtained from equation \ref{E1}. All of the remaining timesteps after $T(\textbf{X})$ until $T_{max}$ are padded with a dummy rule. Then, a 1-hot vector that represents a parse tree is finally obtained. 

During training, a value $\textbf{z}$ is sampled from $q(\textbf{z}|\textbf{X})$ for computing the ELBO

\begin{equation}
    \mathcal{L}(\phi,\theta,\textbf{X})=\mathbb{E}_{q(\textbf{z}|\textbf{X})}[\mathrm{log} \frac{p_\theta(\textbf{X},\textbf{z})}{q_\phi(\textbf{z},\textbf{X})}]
    \label{E3}
\end{equation}

where ${q}(\textbf{z},\textbf{X})$ is a Gaussian distribution. The mean and variance of ${q}(\textbf{z},\textbf{X})$  are the output of the encoder, with an isotropic normal prior $p(\textbf{z})=\mathcal{N}(0, \mathrm {I})$. 

We follow the paper~\cite{kusner2017grammar} for applying non-centered parameterization in equation \ref{E3}.

\subsection{Optimization criteria for RNA design}

To generate RNA sequences for some applications, we need to define specific constraints to optimize the generated RNA sequences. The set of nucleotides is as $B:=$\{A, C, G, U\}. The constraints are as follows:
\begin{itemize}
    \item \textbf{Minimum Free Energy (MFE) Constraint:} The aim is to minimize MFE or put a specific range for MFE.
    \item \textbf{GC-content Constraint:} GC-content which is the percentage of the guanine (G) or cytosine (C) nitrogenous bases in an RNA sequence must be a predefined specific value.
    \item \textbf{Length Constraint (CL):} The length of the sequence must be in a specific predefined range.
    \item \textbf{Secondary structure constraint (CT):} Considering a specific target secondary structure, the aim is to generate a sequence that has that specific target secondary structure.
    \item \textbf{Base positional constraint (CB):} A part of the positions for the generated sequences are restricted for some specific nucleotides based on IUPAC symbols.
    \item \textbf{Mandatory motifs constraint (CM):} All of the generated sequences must contain a predefined group \textit{M} of sequences at any position more than or equal to one time.
    \item \textbf{Forbidden motifs constraint (CF):} A predefined group \textit{F} of forbidden sequences should not exist in any generated sequence.
\end{itemize}

\subsection{Optimization in the latent space of the RGVAE}

For generating RNA sequences based on specific constraints, we need to do optimization in the latent space. In this paper, similar to in~\cite{kusner2017grammar}, Bayesian optimization is used for finding a solution for the problem $\textbf{x}^*= \arg \max_{\textbf{x} \in A} f\left(\textbf{x}\right)$ where $A$ is the known design space as it is stated in~\cite{shahriari2015taking}. Bayesian Optimization is mostly useful for problems where computing $f\left(\textbf{x}\right)$ is computationally costly.  It is a sequential method that uses a prescribed prior over the possible objective functions.  Then the prior is updated to form the Bayesian posterior which is used to induce acquisition functions that guide the exploration to specify the next query point. The details of the Bayesian optimization are described in~\cite{shahriari2015taking}.

In this paper, because finding all evaluations of the features and constraints of the generated RNA sequences is computationally expensive, we use the Bayesian optimization to do fewer evaluations than all of the evaluations.

\section{Results and Discussion}
In this section, we explain the advantages of using the proposed method for generating RNA sequences based on the constraints we already mentioned. We show that using the RGVAE and Bayesian optimization, we optimize the constraints to improve the features. We trained the model based on 101754 tRNA sequences obtained from the Rfam database \cite{CiteDrive2022}. As we will show in the next subsections, We tested two different grammars and realized that one of them outperforms the training data, but the other one does not outperform the training data. So, the performance highly depends on the selected grammar. We examined the dimensions of 10, 25, and 56 for the latent space. The results for the validation data and the optimization for the latent space dimension of 10 is better than 25 and 56.  Therefore, we selected the dimension of the latent space as ten. A three-fold cross-validation is done and different hyper-parameter are evaluated to select the best hyper-parameter. We use the summation of the scores of different constraints for Bayesian optimization. For each scenario, 10 different optimizations with 10 distinct random seeds are done. We did several experiments based on different combinations of constraints. For the optimization of multiple constraints together, we summed the scores of different constraints together to make the objective function. Then, we optimized the objective function using Bayesian optimization. We experimented realistic applications for desiging a riboswitch as well in Section 3.7. 

%To bias the optimization toward each of the scores of different constraints, we used a coefficient for each of the scores and adjusted the coefficients by testing the optimization results for different coefficients. The data for the optimization is generated using the generative model. As much as a coefficient for a score would be larger comparing to other coefficients, the optimization would be more biased toward the corresponding constraint. We did several experiments based on different combinations of constraints.  We select different coefficients for each constraint. For example, we may select one coefficient as one and the second coefficient as 25 for a combination of two constraints. Then, we check the results to see how much the coefficients are close to the desired values. Then, based on the priority of the constraints, we change the coefficients so that the constraint with a higher priority satisfies the desired values. For instance, for a combination of MFE and GC content, if the priority of the optimization is with the GC content to be in the range of $\left[0.48, 0.52\right]$, we increase the coefficients of the GC content score, so that the GC content remains in the desired range.  We experimented realistic applications as well in Section 3.6. 

In addition to the results of our proposed model, we generated RNA sequences randomly based on the uniform distribution, and based on the same grammar of the RGVAE and the same probabilities as the SCFG which was used for training the model. Then, we selected the best RNA sequences based on their features to compare the results with our proposed RGVAE. We will show that the proposed RGVAE outperforms the randomly generated RNA sequences.

The details for each of the experiments and the visualization of the latent space are provided below.

% \subsection{Visualizing the latent space}

% Figure \ref{F3} shows the kernel density estimation (KDE) for all of the ten dimensions of the latent space for all of the encoded training data points. In this figure, the mean and the standard deviation for each of the dimensions are a bit different, the KDE for all of the dimensions is normal which was performed by the RGVAE. 

% \begin{figure}
% \setlength{\fBoxsep}{0pt}%
% \setlength{\fBoxrule}{0pt}%
% \begin{center}
% \includegraphics[scale=0.25]{Kernel.eps}
% \end{center}
% \caption{Kernel Density Estimation (KDE) for ten different dimensions of the latent space.\label{F3}}
% \end{figure}

\subsection{RNA design with MFE minimization for two different grammars}

In this subsection, we evaluate the results of the optimization of the MFE for the generated sequences for two different grammars for the last iteration which was the fifth iteration of optimization. The first grammar we used is the grammar $G$. The second grammar is the grammar $G_0$ as follows:

$G_0: S \rightarrow S|S
$

$
     S \rightarrow aS\hat{a}
$

\begin{figure}
\begin{center}
\includegraphics[scale=0.6]{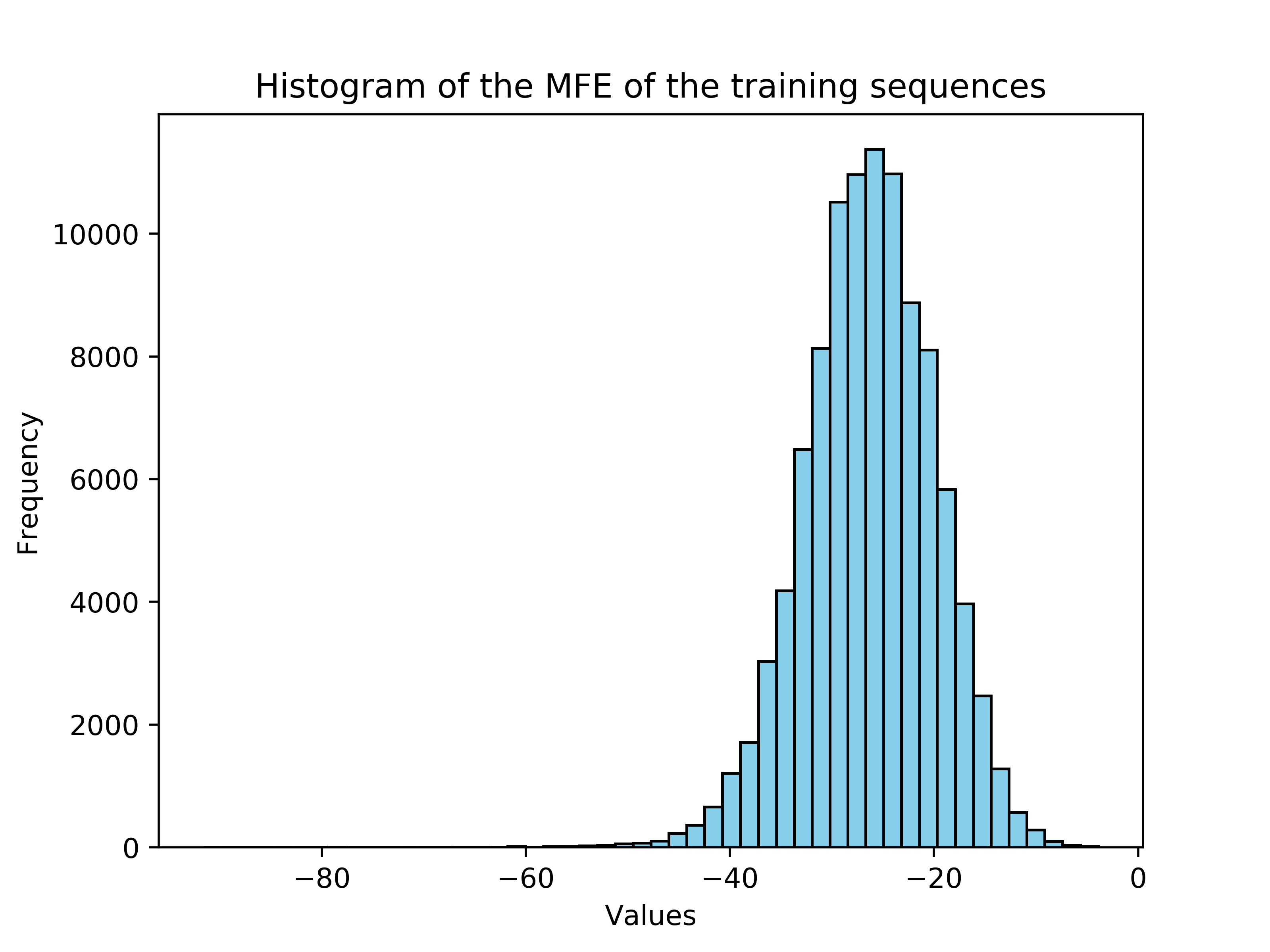}
\end{center}
\caption{Histogram of the MFE of the RNA sequences in the training dataset. \label{F4}}

\end{figure}

\begin{figure}
\begin{center}
\includegraphics[scale=0.5]{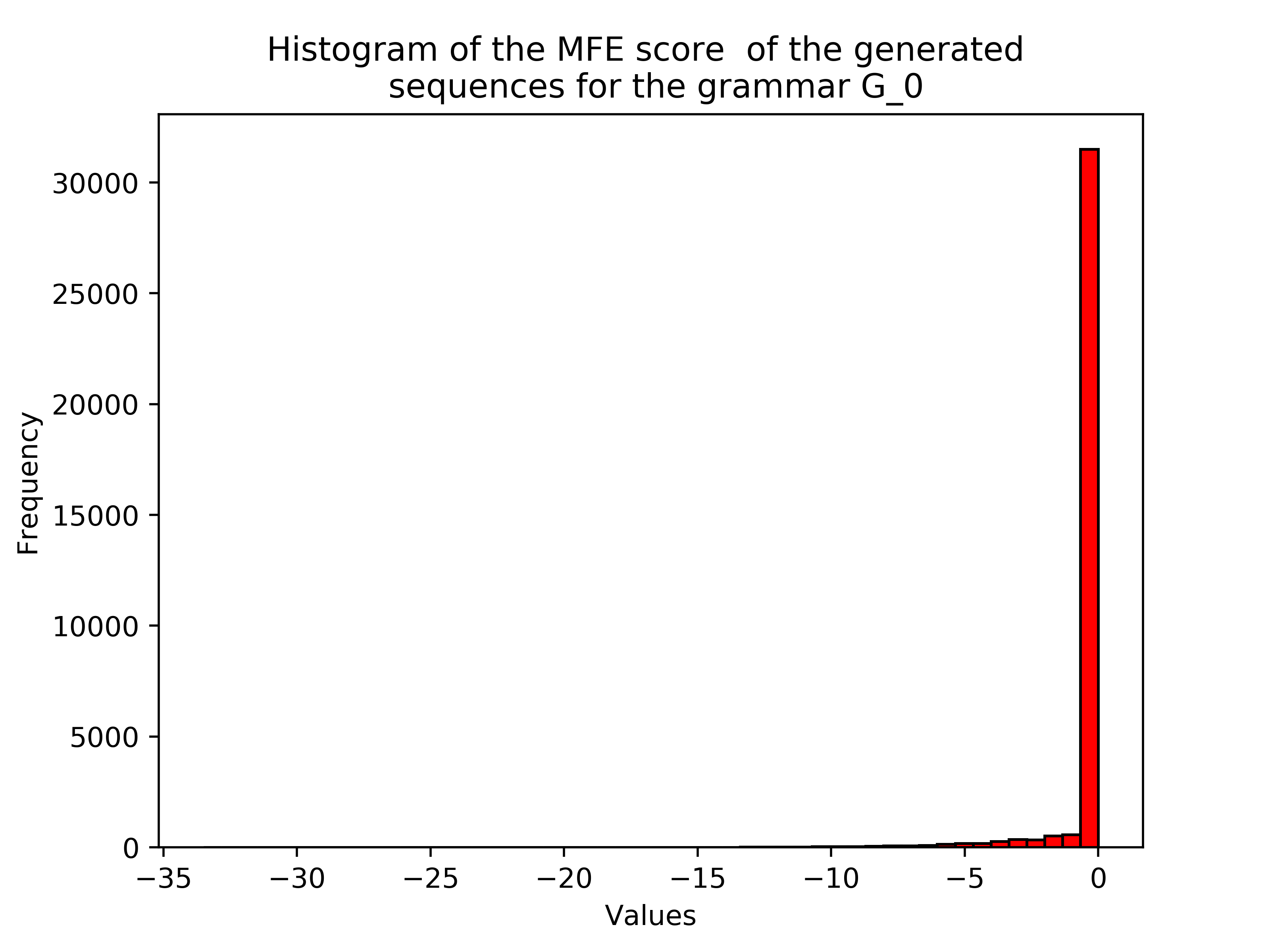}
\end{center}
\caption{Histogram of the MFE of the generated RNA sequences using $G_0$ grammar. \label{F00}}

\end{figure}

\begin{figure}
\begin{center}
\includegraphics[scale=0.5]{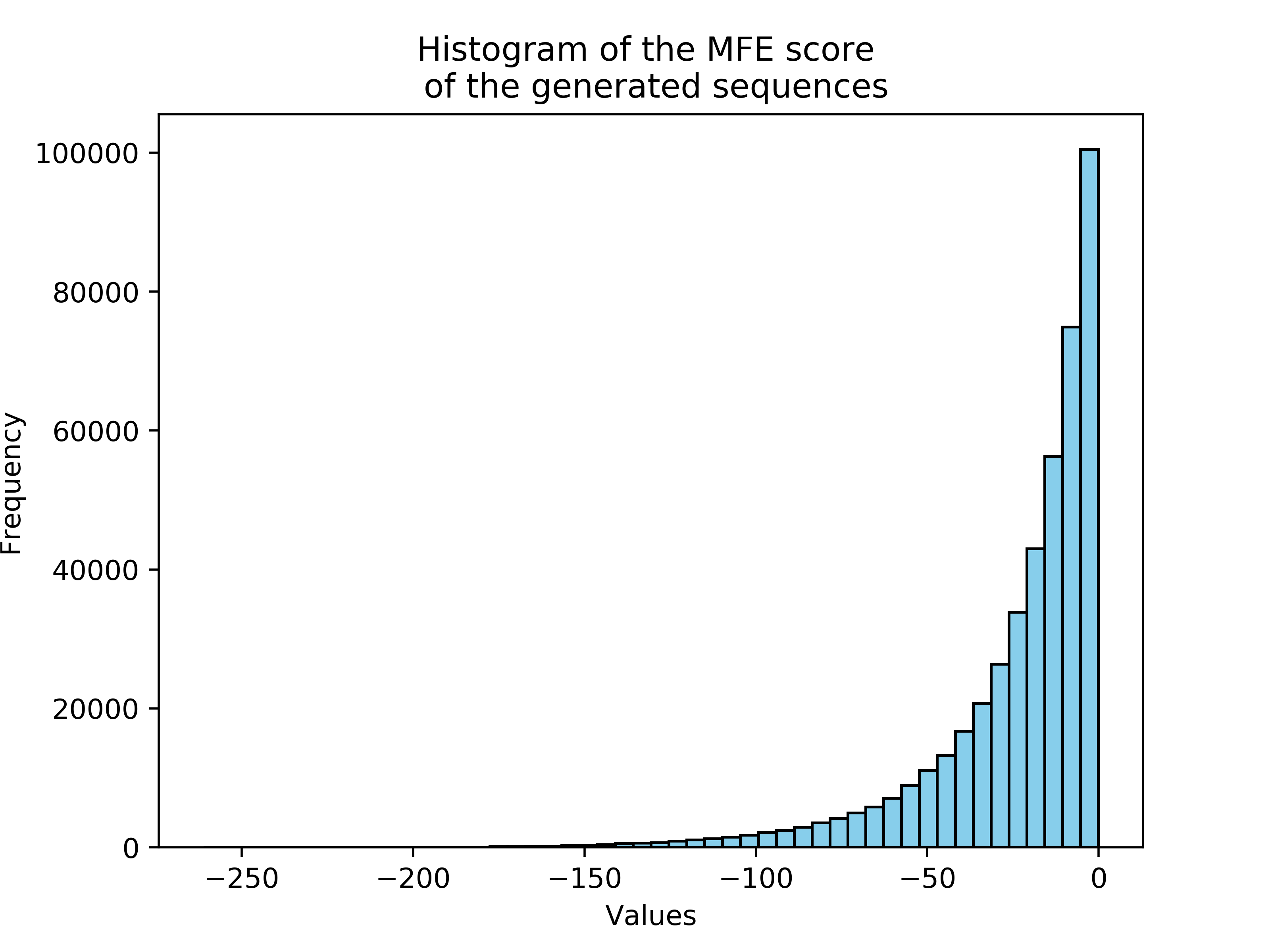}
\end{center}
\caption{Histogram of the MFE of the generated RNA sequences using the grammar $G$. \label{F11}}

\end{figure}

$
     S \rightarrow aS|a
$

where $G0$ is the grammar name, $S$ is the start symbol, $a$ shows any single terminal, and $\hat{a}$ shows the basepair of $a$. Then, we compare the histogram of the generated sequences with the training sequences. The histogram for the MFE of the training RNA sequences is shown in Figure \ref{F4}.  The histograms of the MFE for the grammars $G_0$ and $G$  are shown in Figures \ref{F00} and \ref{F11}, respectively. We removed the generated null sequences and the single nucleotide sequences for plotting the histograms of the generated sequences. By comparing Figure \ref{F11}, \ref{F00}, and \ref{F4} together, we see that the tale of the histogram in Figure \ref{F11} is longer than the Figures \ref{F00} and \ref{F4}. But the tale of the histogram in Figure \ref{F00} is smaller than the Figures \ref{F11} and \ref{F4}. So, by doing the optimization, we generated sequences with lower MFE using the grammar $G$, but the MFE of the optimized generated sequences using the grammar $G_0$ is worse than the training sequences. Therefore, the selection of the grammar highly affects on the performance of the generative model.

\begin{figure}
\begin{center}
\includegraphics[scale=0.5]{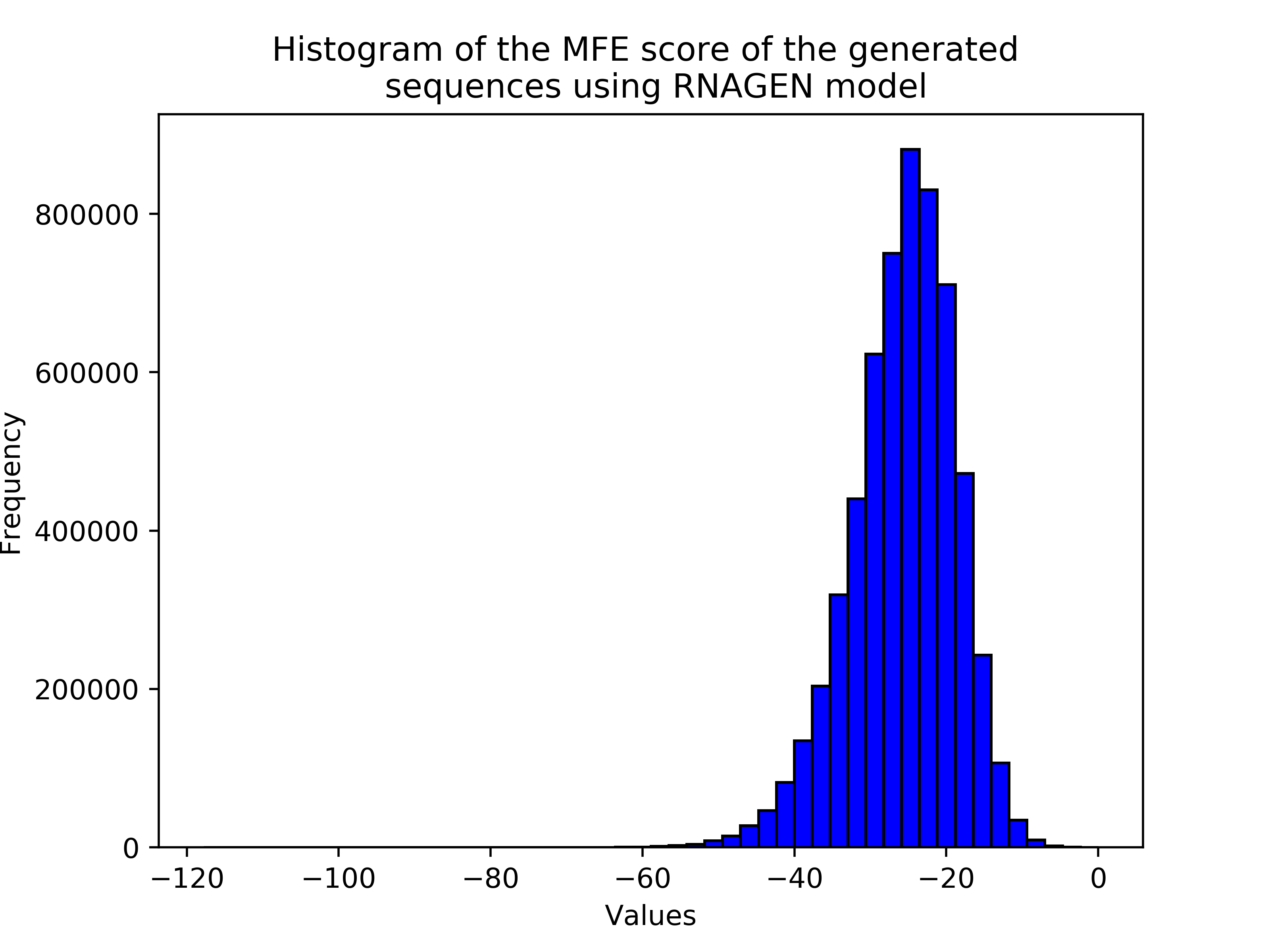}
\end{center}
\caption{Histogram of the MFE of the generated RNA sequences using the RNAGEN model in ~\cite{ozden2023rnagen}. \label{F21}}

\end{figure}

In order to evaluate the performance of the optimization using our proposed RGVAE, we trained the RNAGEN model in ~\cite{ozden2023rnagen} based on the same tRNA data. We generated RNA sequences using this model. Then, we plotted the histogram of the MFE for the generated sequences from the model proposed in ~\cite{ozden2023rnagen}. This histogram is shown in Figure \ref{F21}. By comparing Figures \ref{F11} and \ref{F21}, we realize that our proposed method could generate RNA sequences with lower MFE because the tale of the histogram in Figure \ref{F11} is wider for MFE values less than -70. In addition, The minimum of MFE for the training data is -91.59. The minimum of MFE of the generated RNA sequences using RGVAE and RNAGEN model are -261.2 and -117.80, respectively. So, the minimum of MFE for the generated sequence using RGVAE for grammar $G$ is lower than the minimum MFE of the training data and the generated sequences using RNAGEN model.

\subsection{RNA design with MFE minimization with  target GC-content}

The level of free energy is rarely discussed in RNA design problems although it is of paramount importance in RNA design. In~\cite{esmaili2015erd}, a specific interval for the free energy is considered as a constraint for the designed RNA sequences. In this paper, we aim to minimize the MFE for the generated RNA sequences since the lower amount of MFE usually causes the more stable RNA secondary structures~\cite{wan2012transcriptome, li2011finding, bonnet2004evidence}. 

The GC content is another important factor in generating RNA sequences. The level of GC-content for wild-type sequences in living organisms is usually low or medium, probably for improving the transcription rates and/or structural plasticity~\cite{reinharz2013weighted}. The GC content of a gene region may affect the coverage as well. The 50-60\% GC content in the regions receives the most coverage, but a high (70-80\%) or a low (30-40\%) GC content in a region decreases the coverage very much~\cite{AMR2015251}. In addition, for some structural RNAs, for paired bases, the GC content is considerably higher than the GC content in unpaired bases. Hence, for structured RNA sequences, the secondary structures of the regions with a higher percentage of GC content is probably more stable   ~\cite{chan2009structural}.

%\subsection{RNA design with MFE minimization and  target GC-content}

In this subsection, we provide the results for the generated RNA sequences for different target GC-contents and minimizing MFE.  Then, we compare the results with the training data, random RNA sequences, and RNAGEN model \cite{ozden2023rnagen} to show that the constraints of the generated sequences using RGVAE are improved. Table \ref{Tt0} depicts the best results with the minimum MFE and the GC-content in the range of $\pm 2 \%$ of the target GC-content for different target GC-contents. The results are shown for the training data, RGVAE model, RNAGEN model, and randomly generated RNA sequences. For each target GC-content, we selected the sequences with the minimum MFE in the range of $\pm2\%$ of the target GC-content.  The first column shows the target GC-content of 20\%, 30\%, 50\%, and 70\%. Then, for the training data, and each of generative methods, the second and the third column show the MFE, and the GC-content of the trining and generated sequences, respectively.  N/A in Table \ref{Tt0} means that no sequences found in that range.

\begin{table*}[h]
\small\sf\centering
\begin{center}
\caption{The results for the combination of the MFE and GC-content. For each generating model or the training data, we selected the best sequence with the minimum MFE for the GC-content in the range of $\pm 2\%$   of the target GC-content. \label{Tt0}}

\begin{tabular}{|c|cc|cc|cc|cc|}
\hline
 &\multicolumn{2}{c|}{Training data}&\multicolumn{2}{c|}{RGVAE}&\multicolumn{2}{c|}{RNAGEN}&\multicolumn{2}{c|}{random RNAs} \\
\hline
Target GC-content&MFE&GC-content&MFE&GC-content&MFE&GC-content&MFE&GC-content\\
\hline
20\% &-11.1& 21\%&-39.2& 21\%&N/A& N/A& -41.9 & 20\%\\
30\% &-25.9 & 31\%&-76.3 & 30\%& N/A& N/A&-29.2& 31.6\%\\
50\% &-68.7& 50\%&-215.8& 51\%&-34.09& 50\%&-74.5& 51\%\\
70\% &-65.6& 68\%&-113.2& 68\%&-69.5& 70\%&-77.1& 69.2\%\\

\hline
\end{tabular}\\[10pt]
\end{center}
\end{table*}

%Previous table with random RNAs with grammar.

\begin{figure}
%\setlength{\fBoxsep}{0pt}%
%\setlength{\fBoxrule}{0pt}%
%\begin{center}
\includegraphics[scale=0.5]{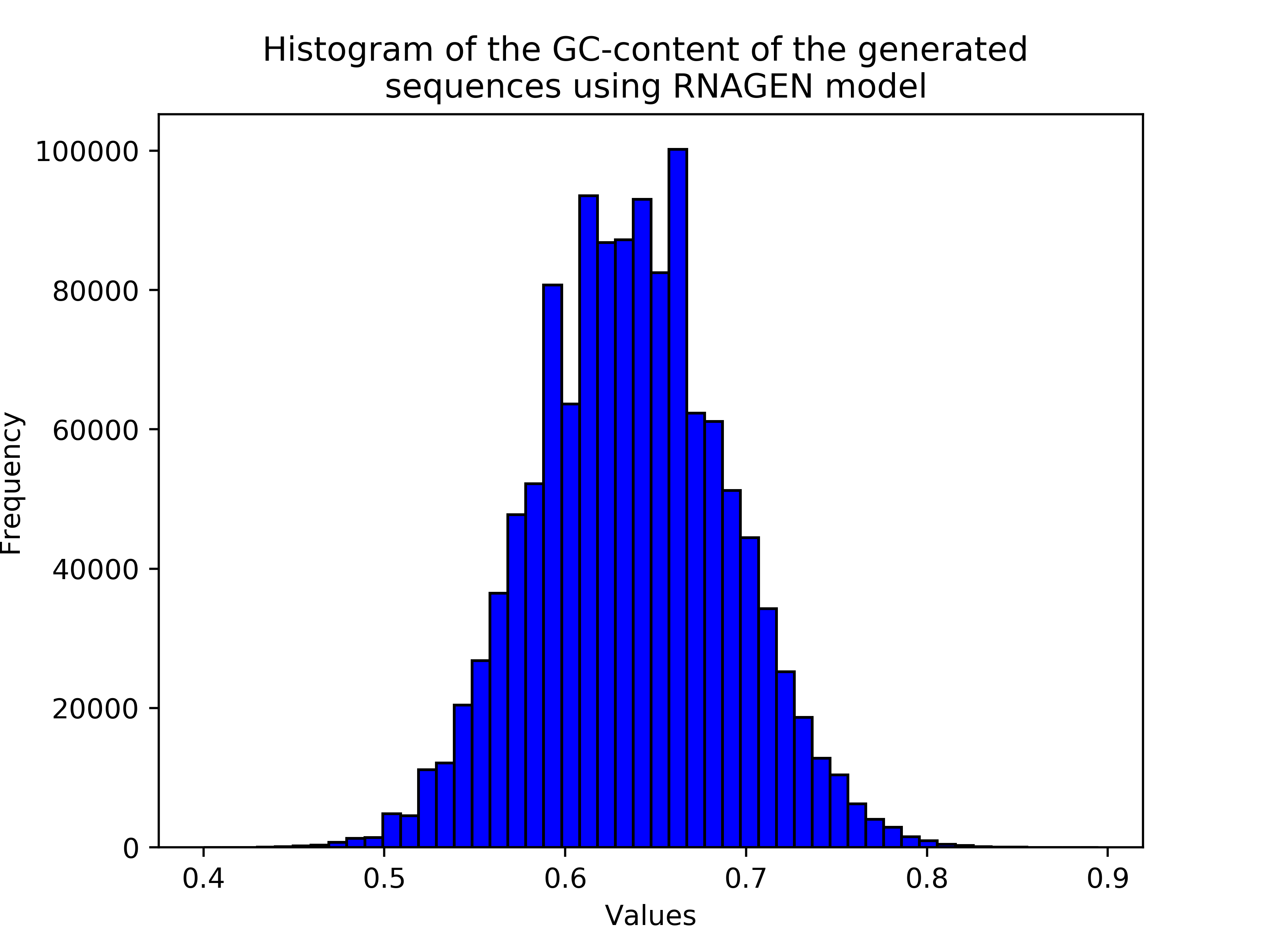}
%\end{center}
\caption{Histogram of the GC-content of the generated RNA sequences using the RNAGEN model in ~\cite{ozden2023rnagen}. \label{F210}}

\end{figure}

\begin{figure}
\begin{center}
\includegraphics[scale=0.6]{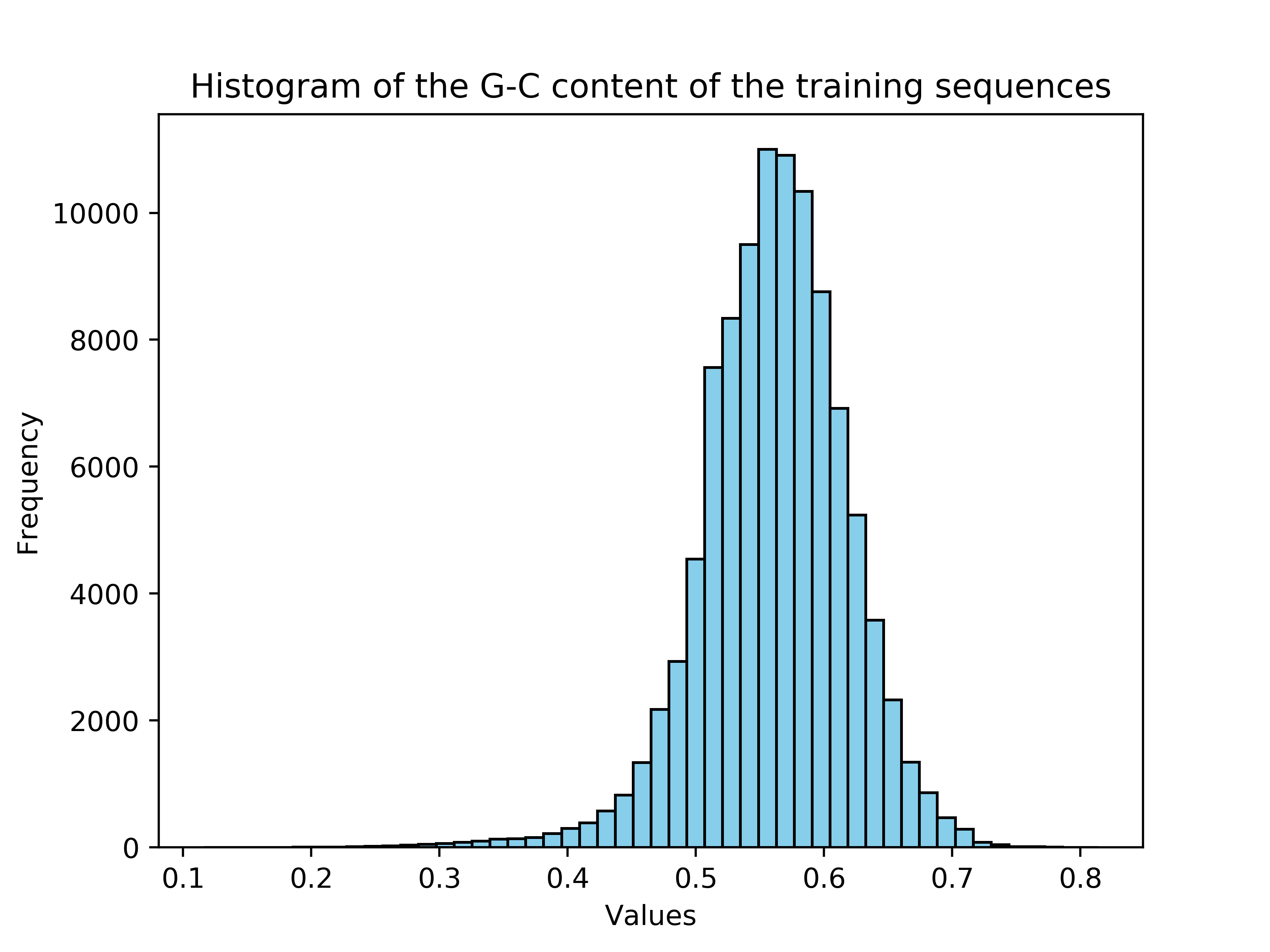}
\end{center}
\caption{Histogram of the GC-content of the RNA sequences in the training dataset \label{F5}}
\end{figure}

\begin{figure}
\begin{center}
\includegraphics[scale=0.6]{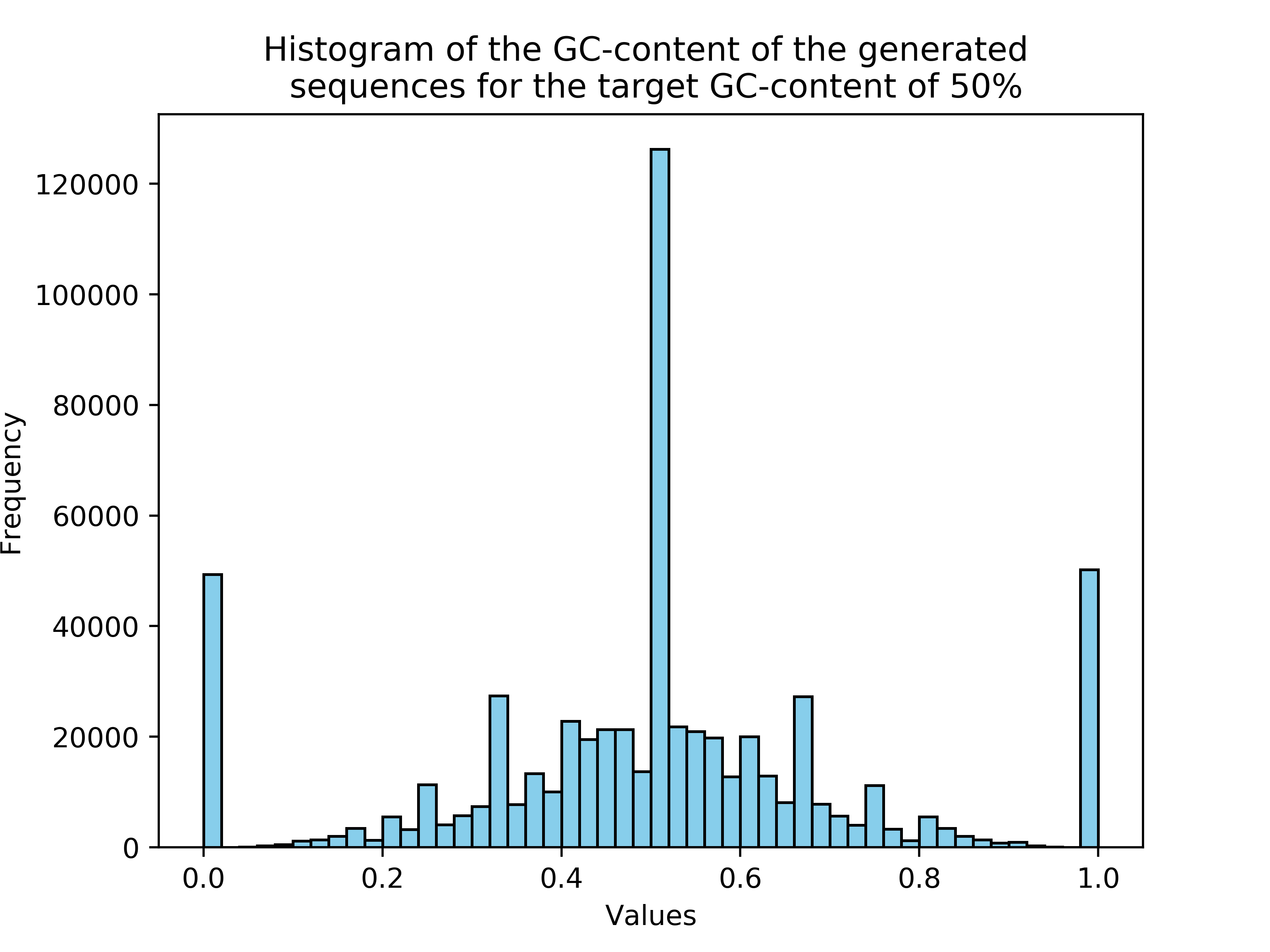}
\end{center}
\caption{Histogram of the GC-content of the generated sequences using RGVAE for the combination of MFE and the target GC-content of 50\%. \label{Fgc50}}
\end{figure}

In order to evaluate the performance of our proposed optimized RGVAE model, we trained the RNAGEN model in \cite{ozden2023rnagen} based on our tRNA data.  The number of the generated sequences for the RNAGEN model, and the random RNAs are the same as RGVAE model. Figures \ref{F210}, and \ref{F5} show the histogram of the GC-content of the generated RNA sequences using the RNAGEN model, and the training data, respectively. Figure \ref{Fgc50} shows the histogram of the generated sequences using RGVAE model for the target GC-content of $50\%$. By comparing the histograms of the GC-contents, we see that for the target GC-content of $50\%$, the peak of the histogram of the RGVAE model, is closer to $50\%$ than the training data and RNAGEN model. As we see in Table \ref{Tt0}, the MFE for the sequences with the GC-content of $20\%$ is higher than the MFE for the sequences with the GC-content of $30\%$, and the MFE for the sequences with the GC-content of $30\%$ is higher than the MFE for the sequences with the GC-content of $50\%$.

In addition, by seeing the results of Table \ref{Tt0}, we see that generally, using our proposed RGVAE model, we find RNA sequences with lower MFE for the same target GC-content compared to the training data, RNAGEN model, and randomly generated RNA sequences without the grammar, except for the target GC-content of $20 \%$ that random RNAs shows the minimum MFE among the other methods in the table. But even for the target GC-content of $20\%$, RGVAE gives a lower MFE than the training data and RNAGEN model. In addition, for the target GC-content of lower than 20\%, and 30\%, RNAGEN model could not generate a sequence.

\begin{figure}
\begin{center}
\includegraphics[scale=0.6]{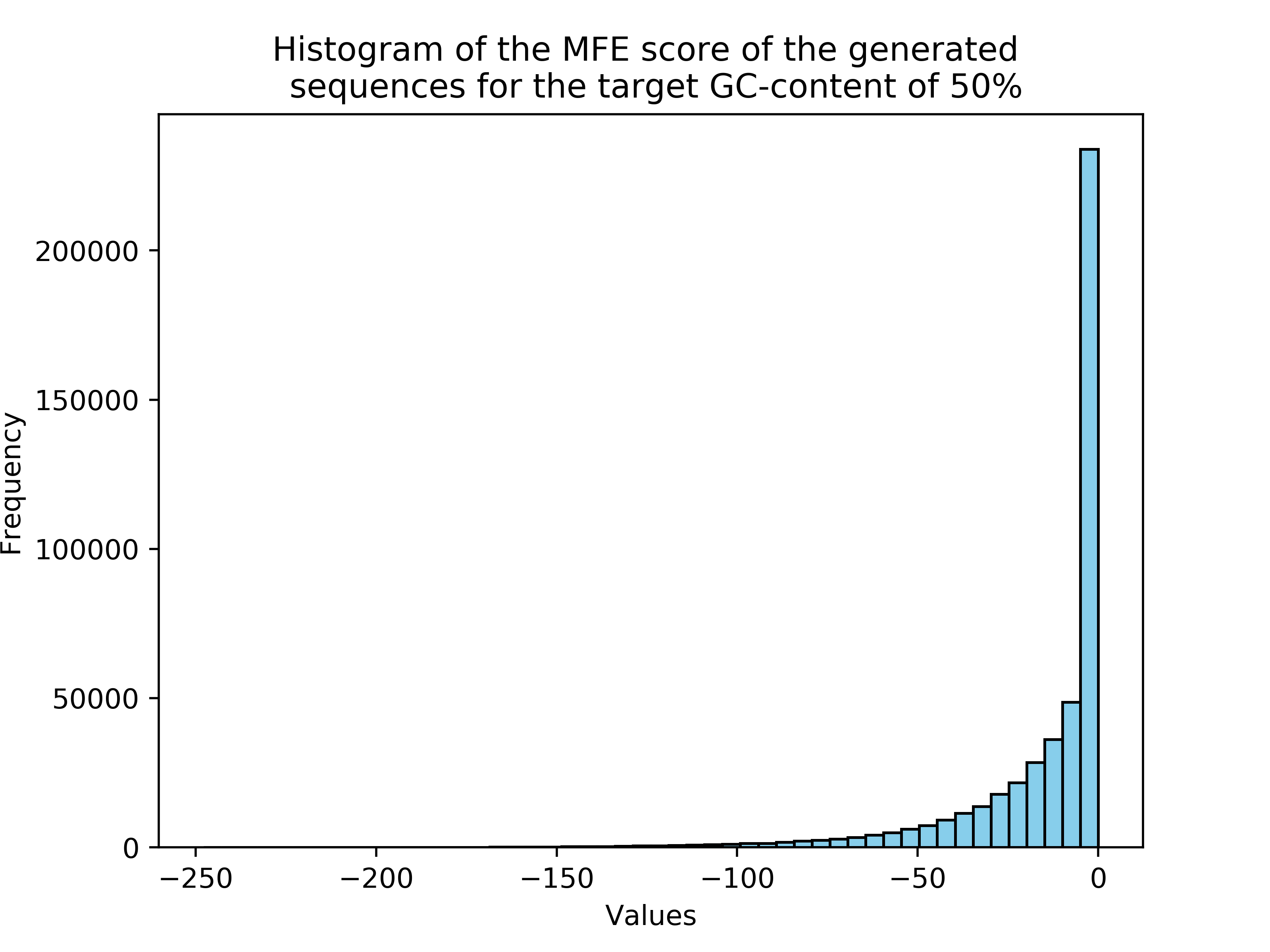}
\end{center}
\caption{Histogram of the MFE of the generated sequences using RGVAE for the combination of MFE and the target GC-content of 50\%. \label{Fmfe50}}
\end{figure}

Figures \ref{F4}, and \ref{F21} show the histogram of the MFE  of the training tRNA data, and the RNAGEN model, respectively.  Figure \ref{Fmfe50} show the histogram of the MFE for the RGVAE model for the target GC-content of $50 \%$. As we see, the tale of the histogram of the MFE for the RGVAE model is wider for the MFE values lower than -60 for the target GC-content of $50 \%$ compared with the histograms of the training data and RNAGEN model.

\subsection{RNA design with MFE minimization, target GC-content, and length constraint}

In this subsection, we investigate the optimization of the combination of the MFE, GC content, and length constraints. We set the target GC content to 50\% and the target length interval between 100 to 150. We selected the sequences with the minimum MFE and the GC-content in the range of 2\% of the target GC-content and the length in the range of the length constraint. The results for the training data and different methods are provided in Table \ref{Ttlen}. As we see the results in this Table, RGVAE generated the best sequence with lower MFE for the target GC-content of 50\% and the length between 100 and 150 comparing to the training data, RNAGEN, random RNAs without grammar and random RNAs with grammar.

\begin{table*}[h]
\small\sf\centering
\begin{center}
\caption{The results for the combination of the MFE, GC-content, and the length constraint. For each generating model or the training data, we selected the best sequence with the minimum MFE for the GC-content in the range of $\pm 2\%$   of the target GC-content of 50\% and the range of the length constraint. \label{Ttlen}}

\begin{tabular}{|c|c|c|c|c|c|}
\hline
 Values&Training data&RGVAE&RNAGEN&random RNAs&random RNAs based on grammar\\
\hline
minimum MFE&-54 &-145.6&N/A&-61.2&-129.5\\
\hline
Target GC-content &49.6\% & 51\%&N/A& 51.3\%&51.3\%\\

\hline
length&
147&149& N/A&146&148 \\

\hline
\end{tabular}\\[10pt]
\end{center}
\end{table*}

%Tables \ref{T8} and \ref{T7} show the results for the combination of the MFE, GC-content, and length constraints of the randomly generated RNA sequences with and without the grammar, respectively. The target GC content is 50\%. The length constraint range is between 100 and 150. The distance of the GC content of the selected sequences is $\pm 5\%$ to the target GC content. Comparing the results in Tables \ref{T8} and \ref{T7} with the results in Table \ref{T3}, it is clear that for the same range of the GC-content and the length, the minimum MFE of our proposed RGVAE is lower than the minimum MFE for the random RNA sequences with and without the grammar. Hence, it shows that RGVAE outperforms the randomly generated RNA sequences with and without grammar.

\begin{figure}
\begin{center}
\includegraphics[scale=0.6]{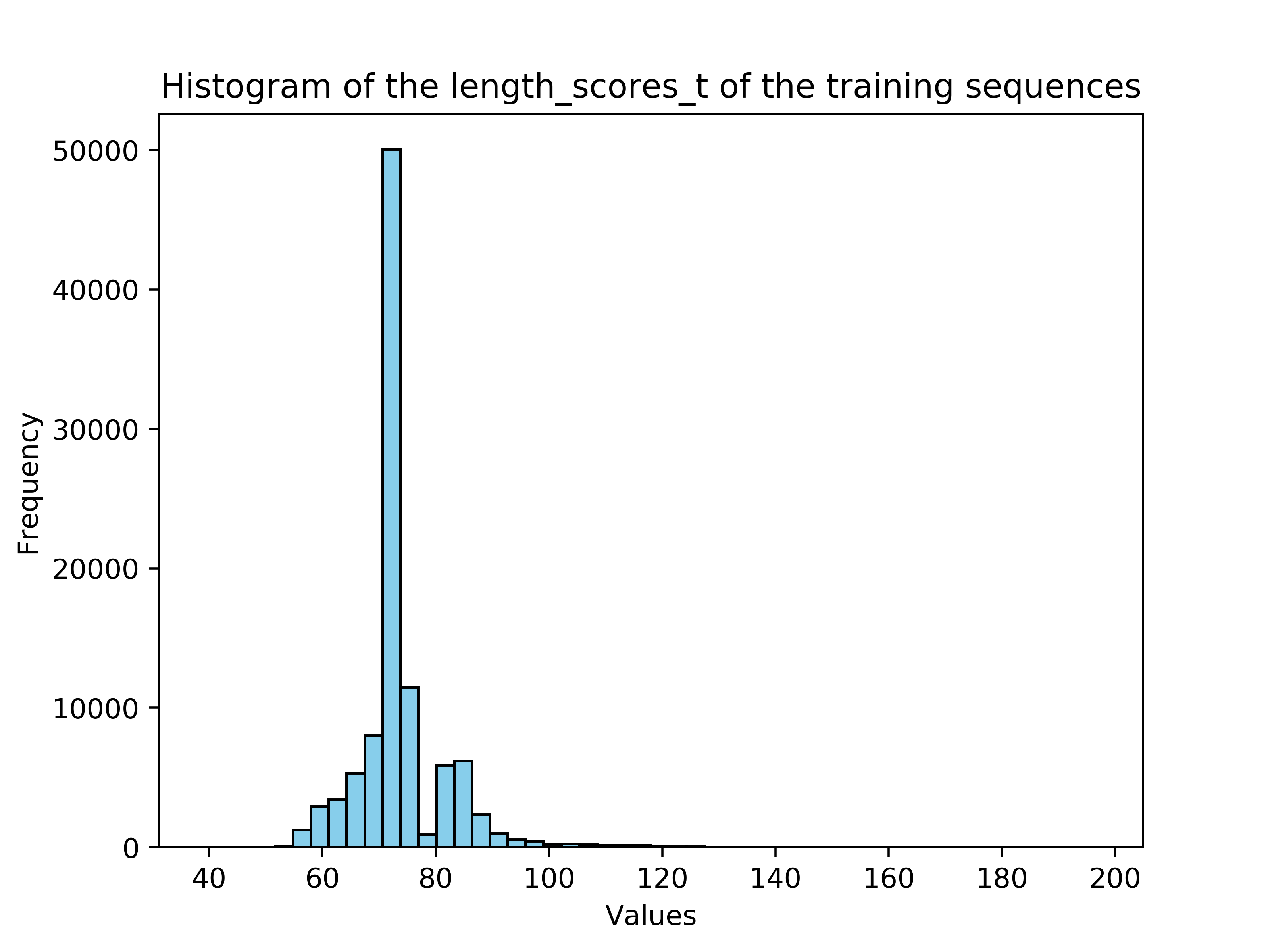}
\end{center}
\caption{Histogram of the length of the training sequences. \label{Ftrain}}
\end{figure}

\begin{figure}
\begin{center}
\includegraphics[scale=0.6]{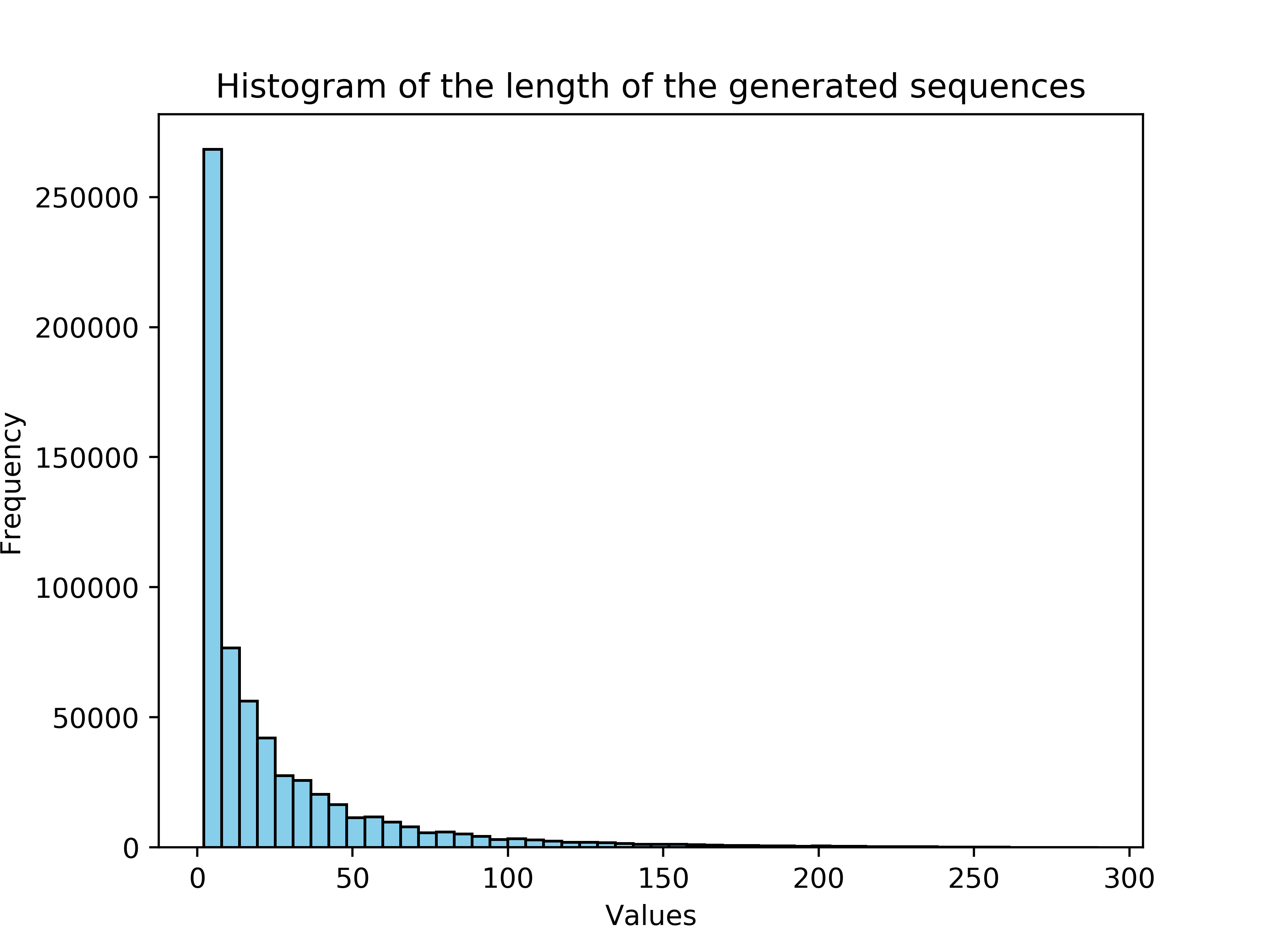}
\end{center}
\caption{Histogram of the length of the generated sequences using RGVAE for the combination of MFE, the target GC-content of 50\%, and the length between 100 and 150. \label{FLenGenRGVAE}}
\end{figure}

\begin{figure}
\begin{center}
\includegraphics[scale=0.6]{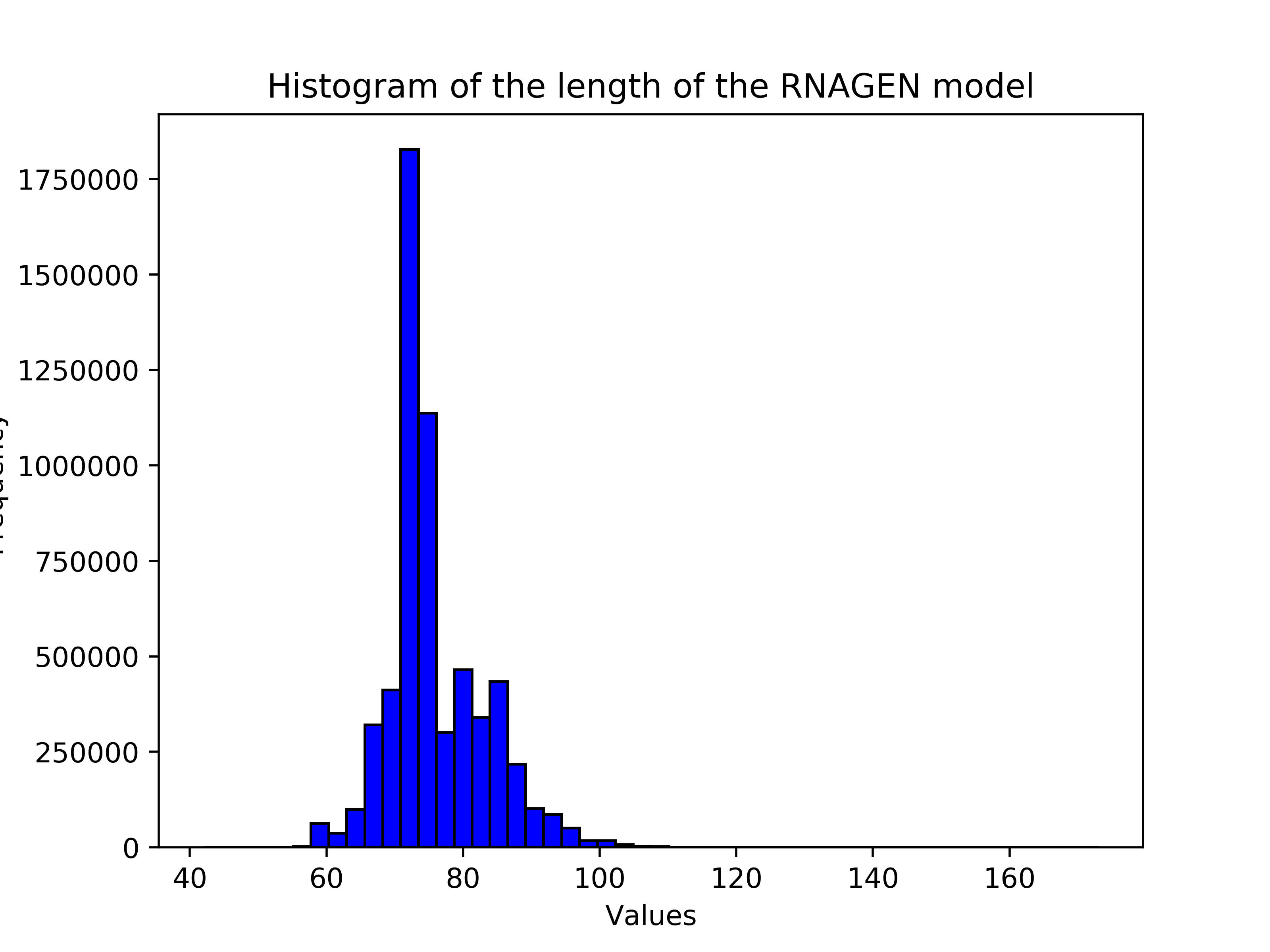}
\end{center}
\caption{Histogram of the length of the generated sequences using RNAGEN for the combination of MFE, the target GC-content of 50\%, and the length between 100 and 150. \label{FLenGenRNAGEN}}
\end{figure}

\begin{figure}
\begin{center}
\includegraphics[scale=0.6]{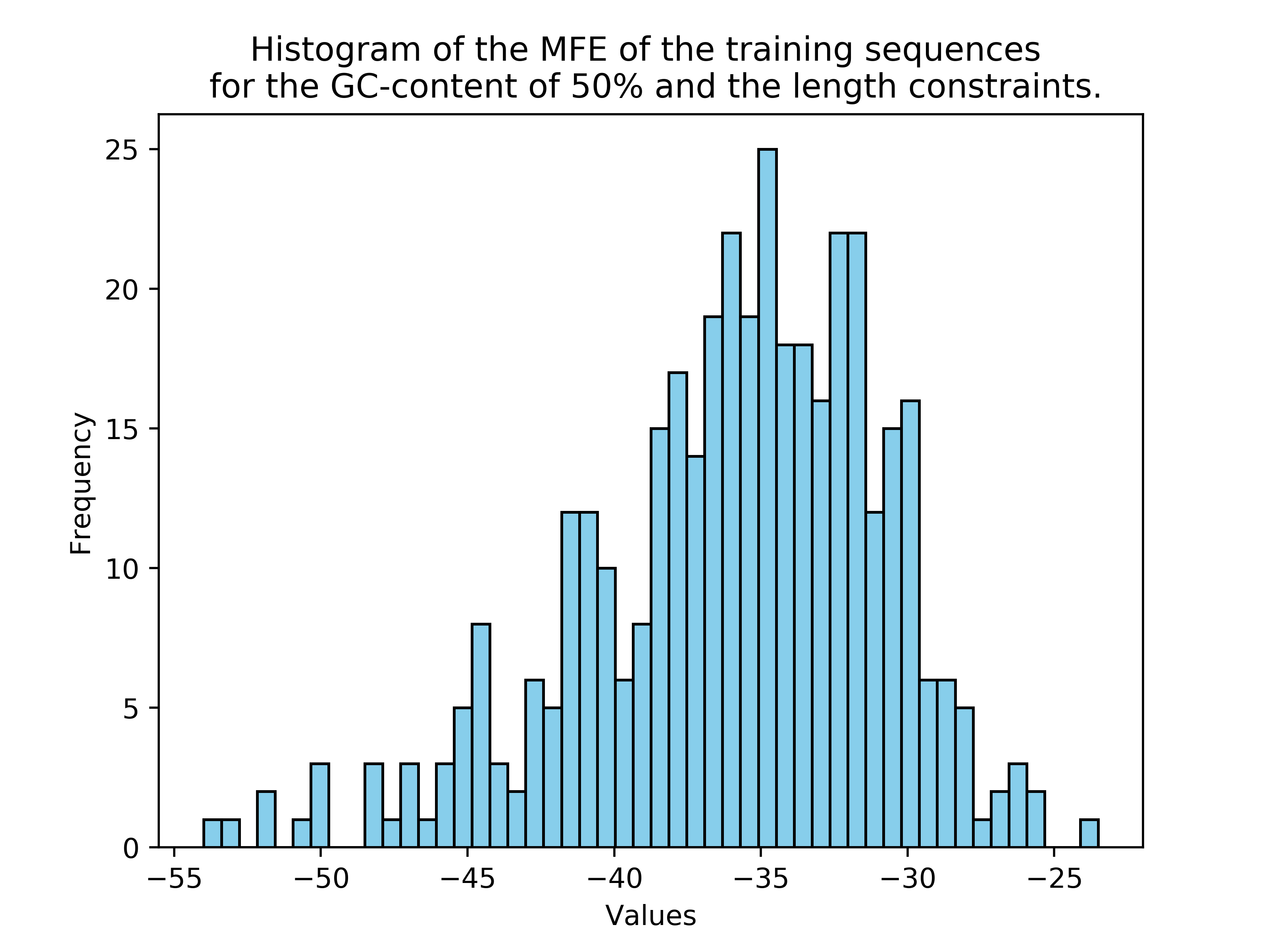}
\end{center}
\caption{Histogram of the MFE of the selected sequences of the training data with the target GC-content of 50\%, and the length between 100 and 150. \label{FTrainJustLen}}
\end{figure}

\begin{figure}
\begin{center}
\includegraphics[scale=0.6]{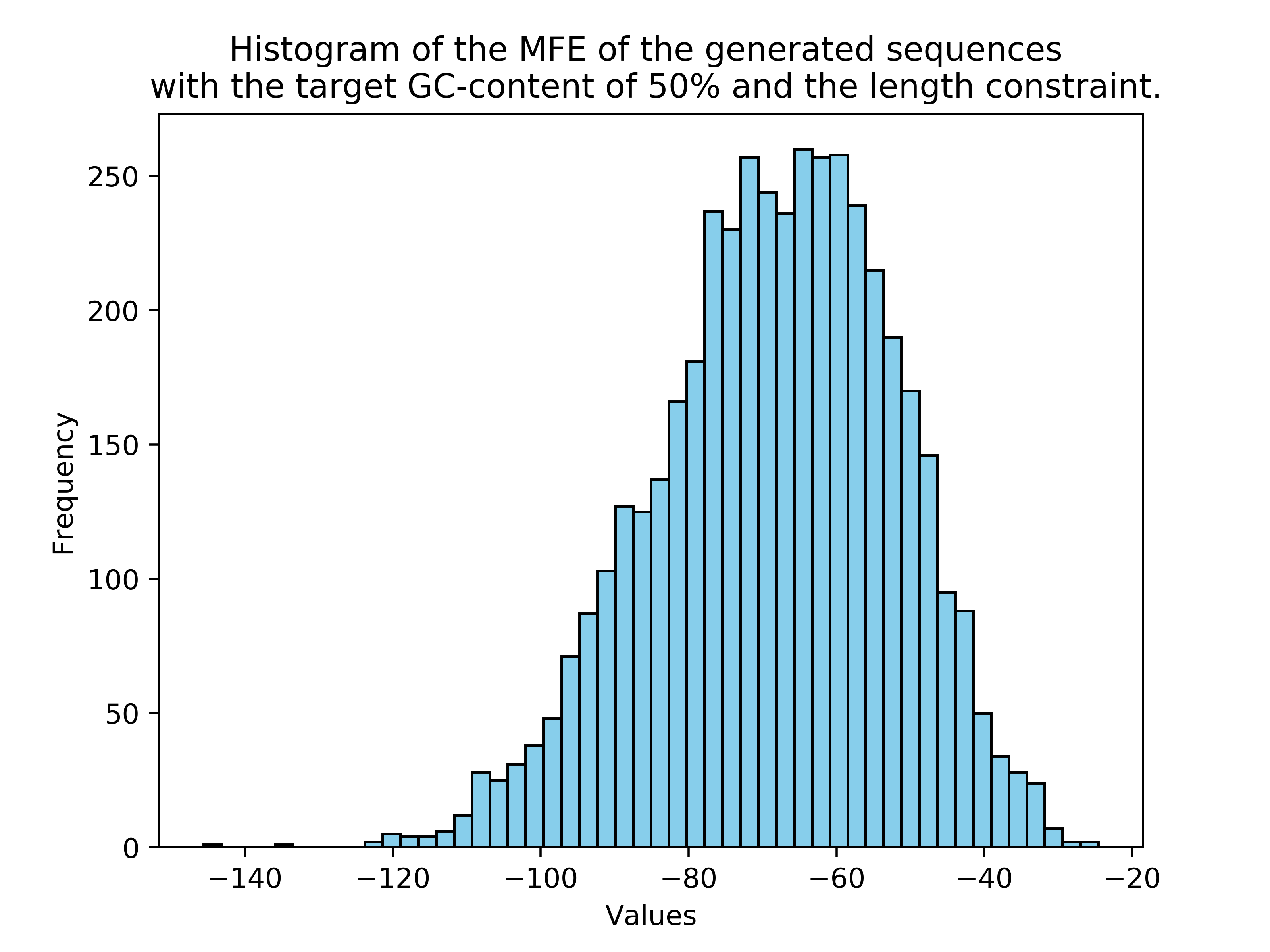}
\end{center}
\caption{Histogram of the MFE of the selected generated sequences using RNAGEN with the target GC-content of 50\%, and the length between 100 and 150. \label{FGenJustLen}}
\end{figure}

The histograms of the length of the training sequences, the generated sequences using RGVAE, and the generated sequences using RNAGEN are shown in Figures \ref{Ftrain}, \ref{FLenGenRGVAE}, and \ref{FLenGenRNAGEN}, respectively. As we see, the tail of the histogram in Figure \ref{FLenGenRGVAE} for the lengths between 100 and 150 is longer than Figures \ref{Ftrain} and \ref{FLenGenRNAGEN}. In addition, the histograms of the MFE of the selected sequences with the GC-content of $\pm 2\%$ of the target GC-content of 50 \% and the length between 100 and 150 for the training data and the generated sequences using RGVAE are shown in Figures \ref{FTrainJustLen} and \ref{FGenJustLen}. As we see in these figures, the peak of the histogram in Figure \ref{FGenJustLen} shows a lower MFE compared with the peak of the histogram in Figure \ref{FTrainJustLen}. Furthermore, the tail of the Figure \ref{FGenJustLen} is wider for lower MFE values than the Figure \ref{FTrainJustLen}. For the RNAGEN model, no sequences found with the selected constraints.

\subsection{RNA design with MFE minimization, target GC-content, mandatory and forbidden motifs}

\begin{figure}
\begin{center}
\includegraphics[scale=0.5]{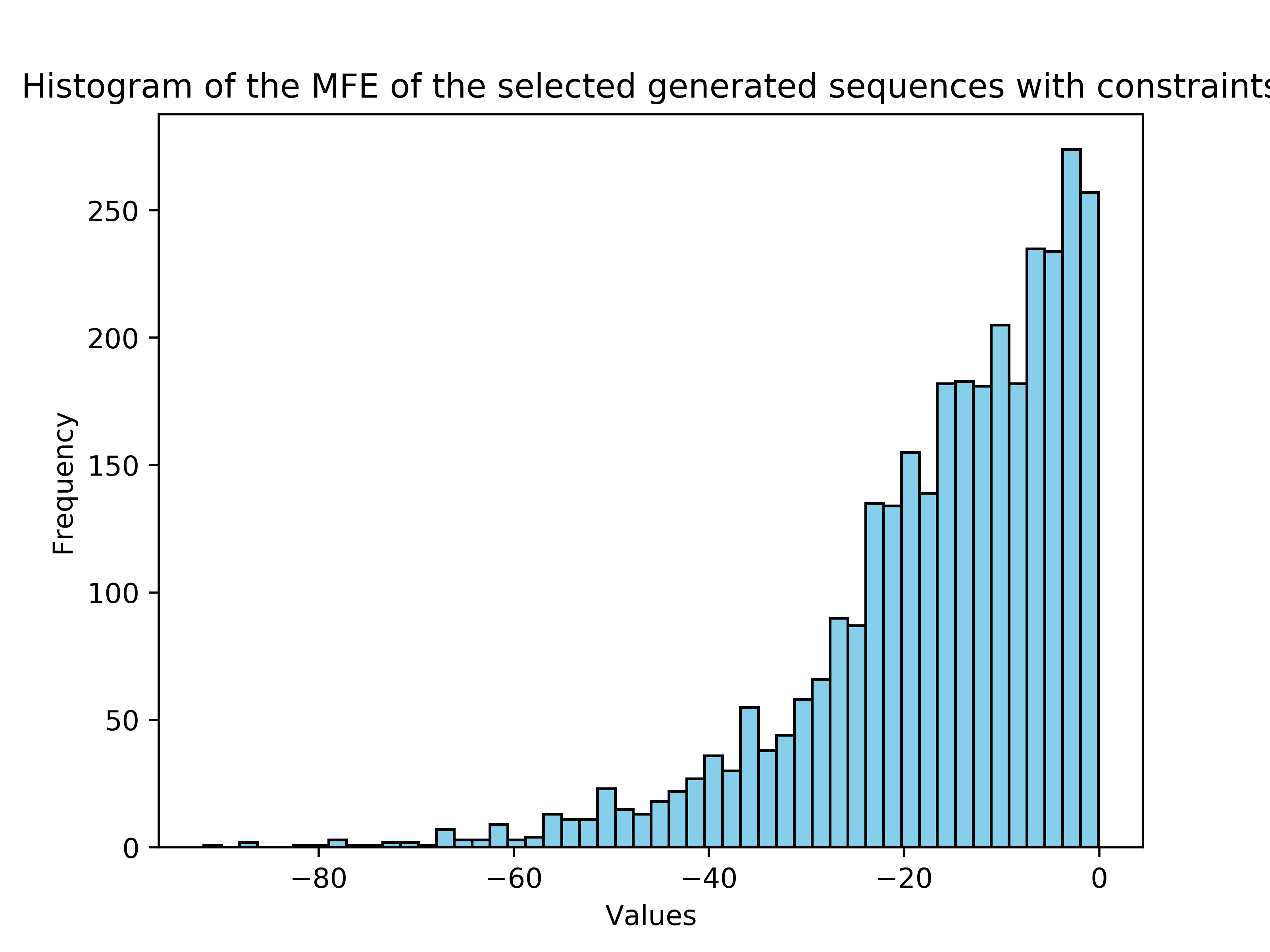}
\end{center}
\caption{Histogram of the MFE of the selected generated sequences using RGVAE with the target GC-content of 50\%, the mandatory motif as M=\{CGU\}, and the set of forbidden motifs as F = \{AAU, CGC, UGC\}. \label{FGenMFJustMFE1}}
\end{figure}

\begin{figure}
\begin{center}
\includegraphics[scale=0.5]{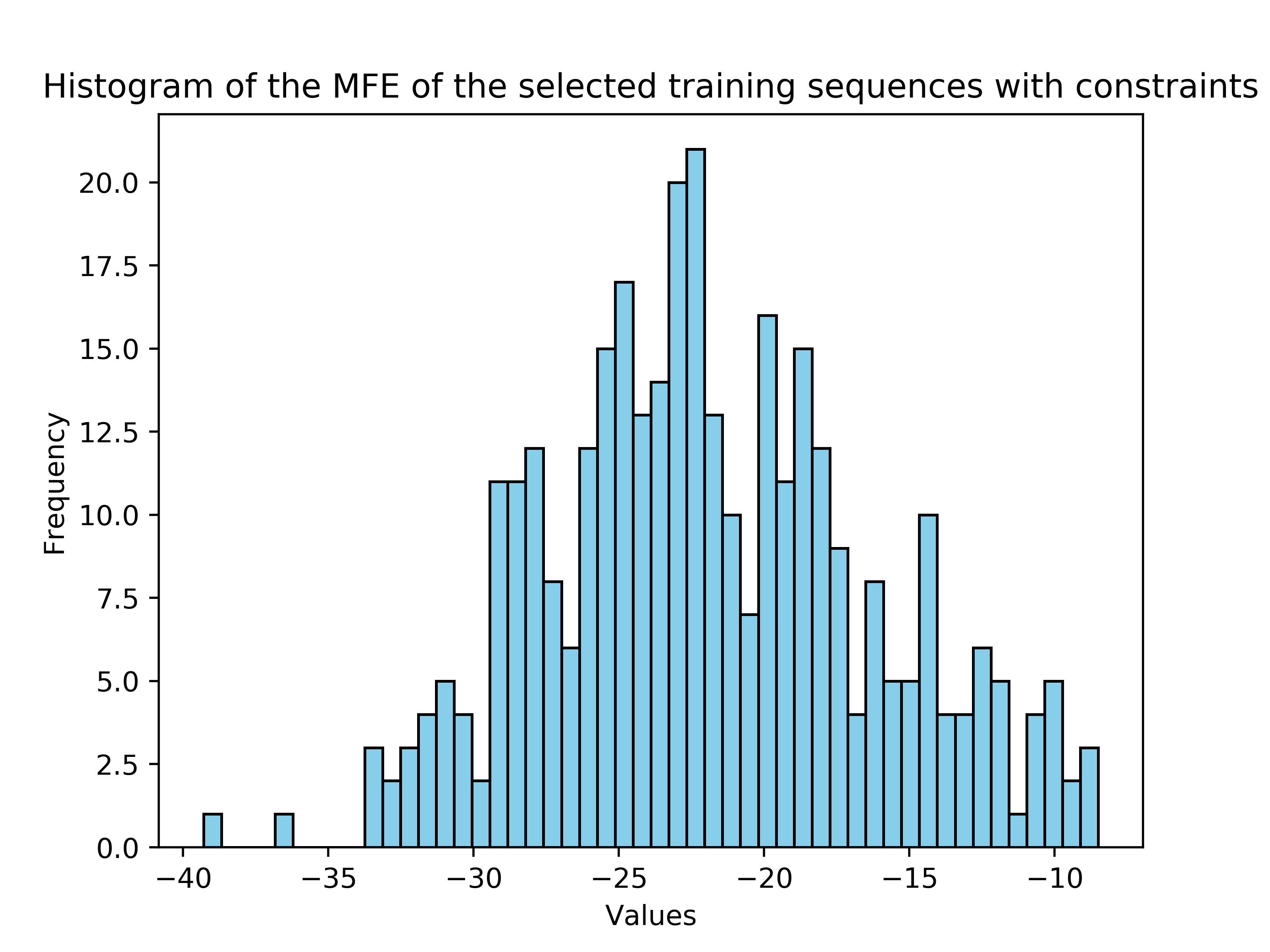}
\end{center}
\caption{Histogram of the MFE of the selected training sequences with the target GC-content of 50\%, the mandatory motif as M=\{CGU\}, and the set of forbidden motifs as F = \{AAU, CGC, UGC\}. \label{FTrainMFJustMFE1}}
\end{figure}

\begin{figure}
\begin{center}
\includegraphics[scale=0.5]{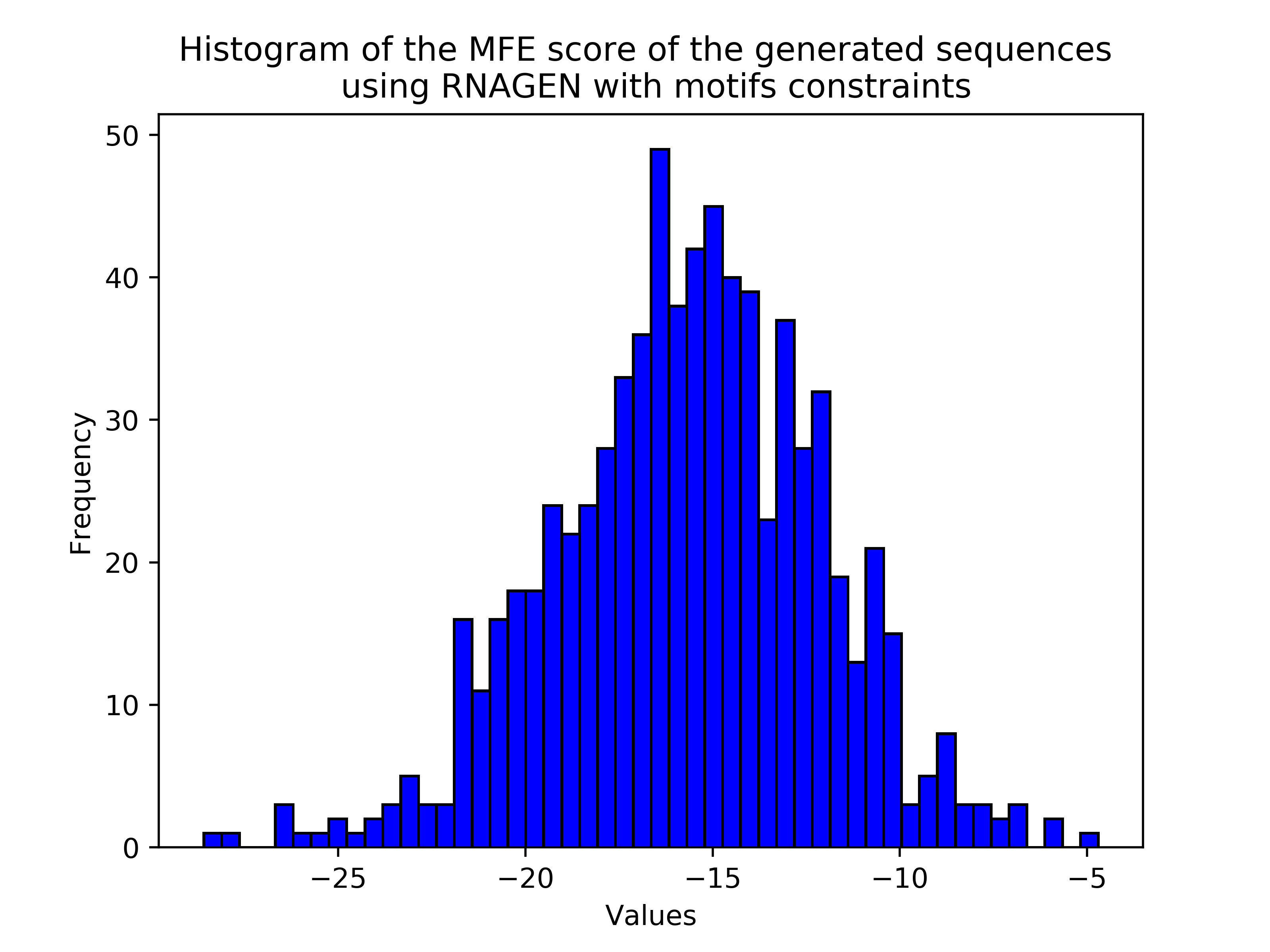}
\end{center}
\caption{Histogram of the MFE of the selected generated sequences using RNAGEN with the target GC-content of 50\%, the mandatory motif as M=\{CGU\}, and the set of forbidden motifs as F = \{AAU, CGC, UGC\}. \label{FMFE_RNAGEN_MF1}}
\end{figure}

In this subsection, we consider the combination of the MFE, GC-content, mandatory, and forbidden motifs for two different experiments. In the first experiment, we consider the set of mandatory motifs as $M=\{CGU\}$ and the set of forbidden motifs as $F=\{AAU, CGC, UGC\}$ as arbitrary motifs which were considered in the paper~\cite{zhou2013flexible}. In~\cite{zhou2013flexible}, these motifs were considered for RNA design problems with a target secondary structure. But we investigate the results of the combination of these motifs with MFE and GC content in this paper. We set the target GC content to 50\%. In this subsection, we selected the best sequence from the generated sequences using RGVAE which has the GC-content of $\pm 2\%$ of the target GC-content, satisfies the mandatory and forbidden motifs, and has the minimum MFE. The generated sequence using RGVAE is CUUCCCGUAG CCCACAACAU UGGAAGCUGG UGUGUACCAC AGGAAACACU UUACCUUUCC UAAAGGGUAU CCGUUGGAUA CCCUUUAAGG AAAGGUAGUG UUUCCUGUGG UAUGUGGGCU ACGGGAAG which satisfies the motifs constraints, has the GC-content of 48.43\%, and the MFE is -91.8\%. The histograms of the MFE of the selected generative sequences using RGVAE, training sequences, and the generated sequences using RNAGEN which satisfies the constraints are shown in Figures \ref{FGenMFJustMFE1}, \ref{FTrainMFJustMFE1}, and \ref{FMFE_RNAGEN_MF1}. As we see in these figures, RGVAE histogram has a longer tail and could generate sequences with lower MFEs comparing to RNAGEN and the training sequences.

In the second experiment, the motifs are selected based on realistic applications. The set of the mandatory motif is $M=\{CCU\}$ which is mentioned in~\cite{mitchell2005identification} and is identified as a part of a  permits internal ribosome entry and polypyrimidine-rich tract. The set of the forbidden motifs is selected as $F=\{GGA, GGG, CUC\}$. These motifs are provided in~\cite{bourdeau1999distribution}, for Rev-Binding Elements (RBE). As it is indicated in~\cite{bourdeau1999distribution}, the transport of unspliced transcripts of the HIV genome is promoted to the cytoplasm by the Rev protein and its binding site, the Rev responsive element (RRE). The target GC content is set to 50\%. The generated RNA sequence is CGUUGUUAAG GCCGAUUUUU GGUCGACUAA GCGCGCGCCA AAUCUAUUUC AAGGUGAGCA GCAUUGACCA CCAAAUUGGU GGUCAAGCUG CUGCAGCGCG CGCUUAGUCG ACCAAAAAUC GGCCUUAACA ACG which satisfies the motifs constraints, has a GC-content of 51.12\% which is close to the target GC-content, and MFE of -100.59.

\begin{figure}
\begin{center}
\includegraphics[scale=0.5]{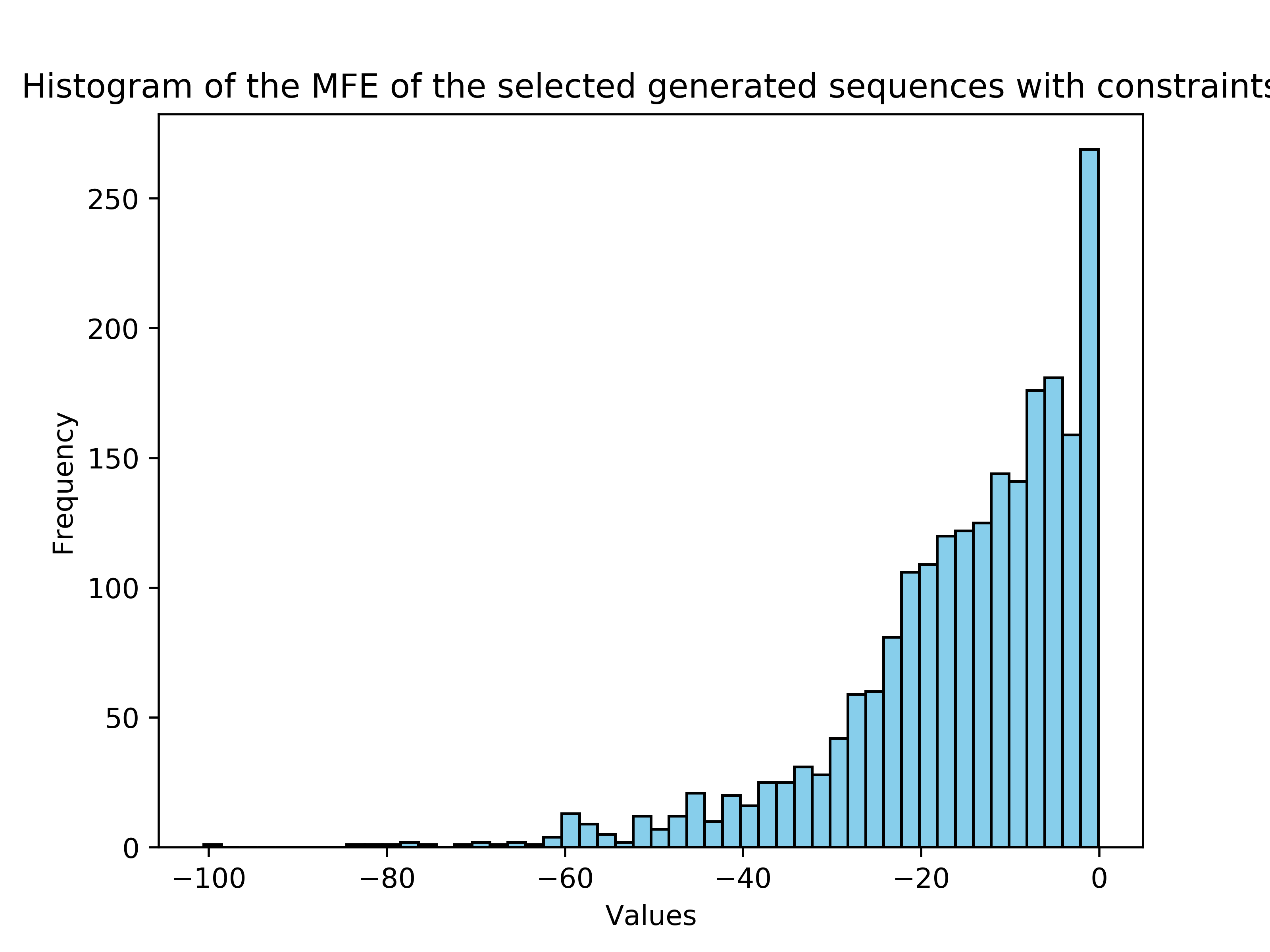}
\end{center}
\caption{Histogram of the MFE of the selected generated sequences using RGVAE with the target GC-content of 50\%, the mandatory motif as M=\{CCU\}, and the set of forbidden motifs as F = \{GGA, GGG, CUC\}. \label{FGenMFJustMFE2}}
\end{figure}

\begin{figure}
\begin{center}
\includegraphics[scale=0.5]{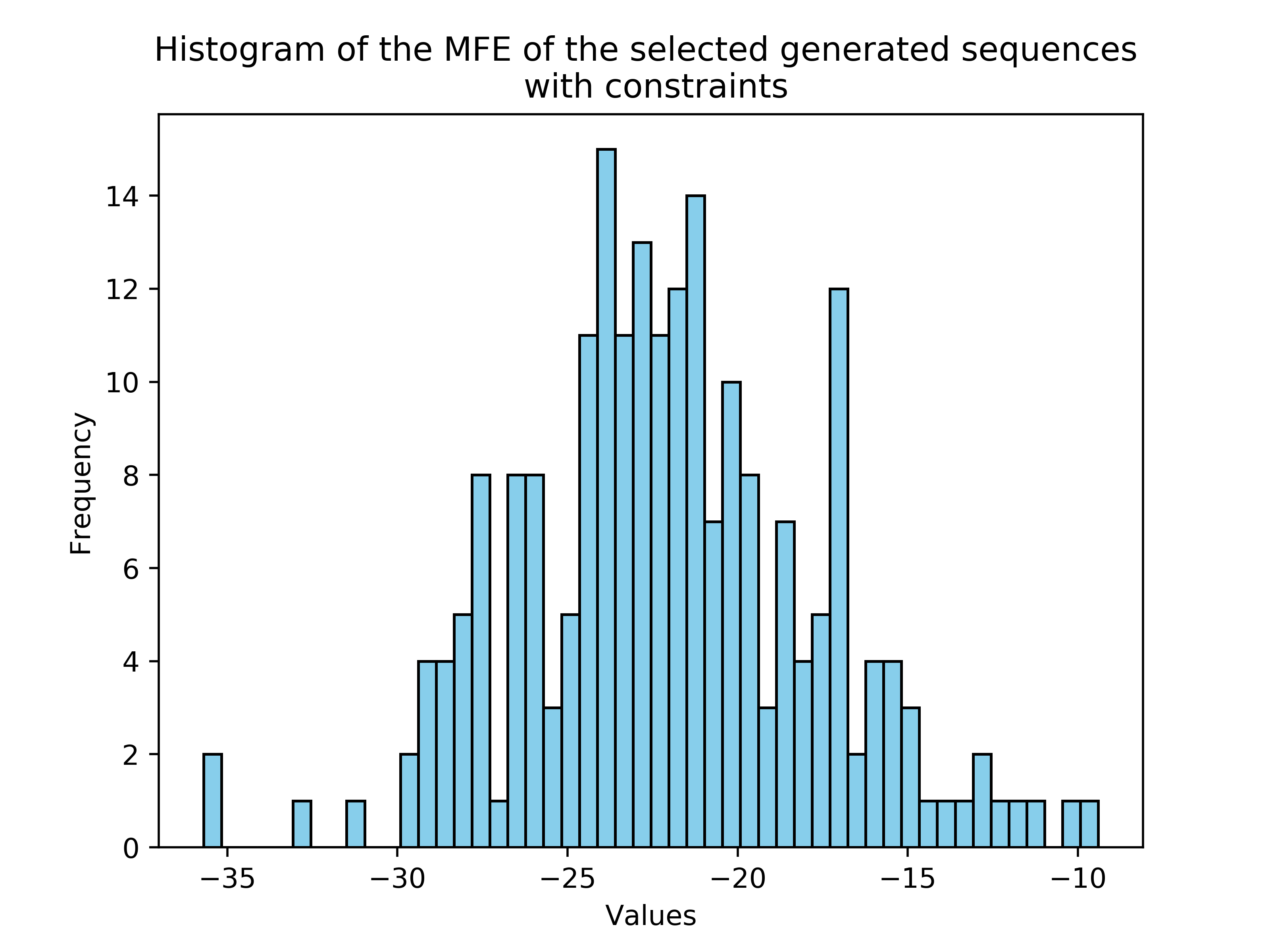}
\end{center}
\caption{Histogram of the MFE of the selected training sequences with the target GC-content of 50\%, the mandatory motif as M=\{CCU\}, and the set of forbidden motifs as F = \{GGA, GGG, CUC\}. \label{FTrainMFJustMFE2}}
\end{figure}

\begin{figure}
\begin{center}
\includegraphics[scale=0.5]{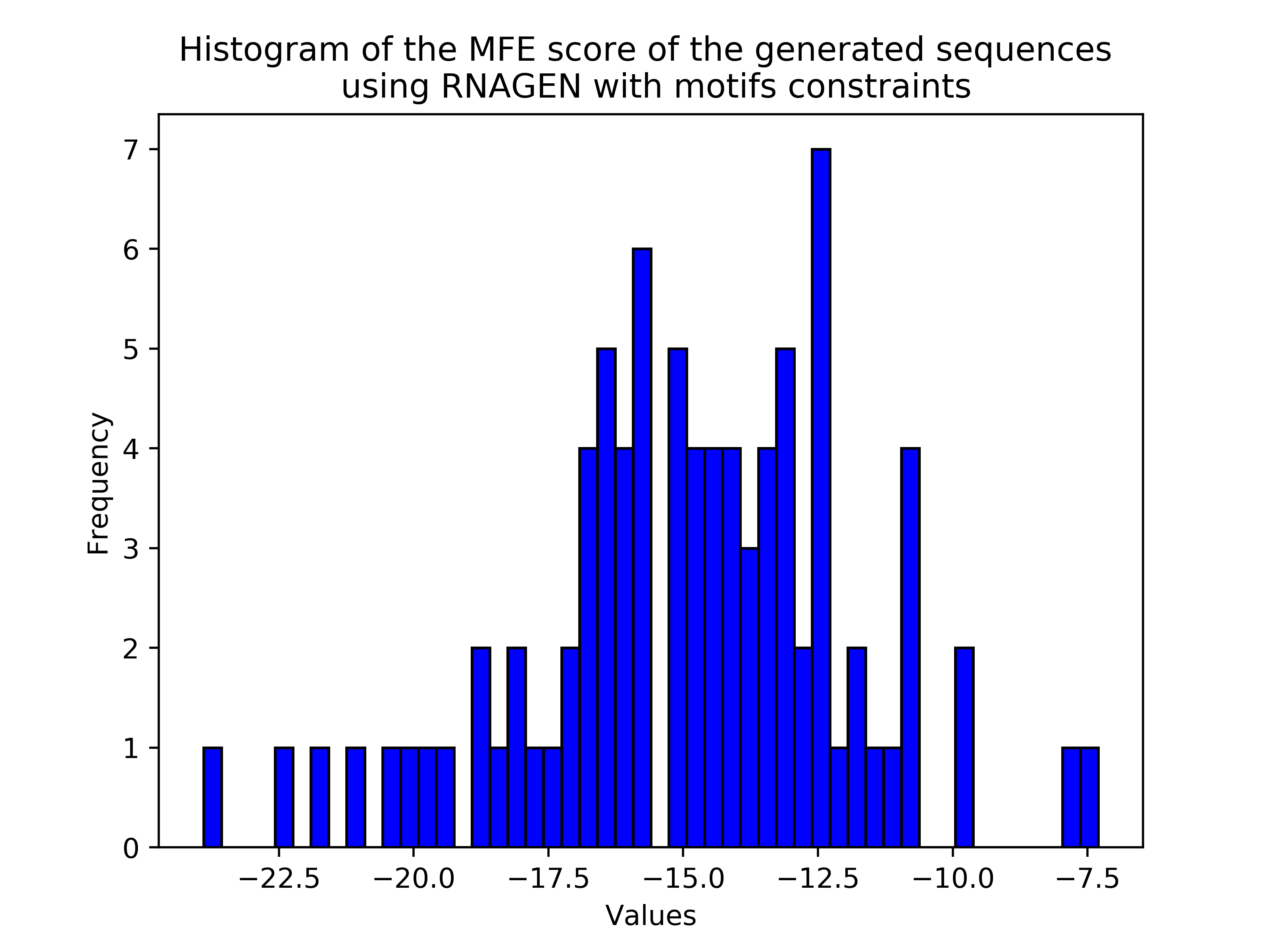}
\end{center}
\caption{Histogram of the MFE of the selected generated sequences using RNAGEN with the target GC-content of 50\%, the mandatory motif as M=\{CCU\}, and the set of forbidden motifs as F = \{GGA, GGG, CUC\}. \label{FMFE_RNAGEN_MF2}}
\end{figure}

The histograms of the MFE of the selected generative sequences using RGVAE, training sequences, and the generated sequences using RNAGEN which satisfies the constraints are shown in Figures \ref{FGenMFJustMFE2}, \ref{FTrainMFJustMFE2}, and \ref{FMFE_RNAGEN_MF2}. As we see in these figures, RGVAE histogram has a longer tail and generates sequences with lower MFEs comparing to RNAGEN and the training sequences.

By considering the results, we conclude that RGVAE could successfully generate the sequences which satisfy the constraints and have lower MFE comparing with the training data, and the RNAGEN model.

\subsection{RNA design with target GC-content, mandatory motifs, and base positional constraint}

For designing an RNA sequence, we consider a specific target secondary structure, and the aim is to find an RNA sequence that has that specific RNA secondary structure~\cite{busch2006info},~\cite{runge2018learning}. Designing RNA secondary structures is NP-hard \cite{bonnet2020designing}. For measuring the distance between the secondary structure of the generated sequence and the target secondary structure, we use the hamming distance as a loss function as it is mentioned in~\cite{busch2006info}. 

In this subsection, we provide the results for the combination of RNA design, GC-content, mandatory motifs, and base positional constraints. We consider "......(((....)))" as the target secondary structure which was used in~\cite{zhou2013flexible}. In~\cite{zhou2013flexible}, the aim was to design an RNA sequence based on the target secondary structure "......(((....)))", mandatory and base positional constraints "UCGUCG". But in this paper, for the mandatory motifs and base positional constraints, we consider the first two nucleotides in the sequence to be "UC". Therefore, the total sequence must be "UCNNNNNNNNNNNNNN" where N can be any of the nucleotides in \{A, U, C, G\}. In addition, we have a constraint on GC-content which was not considered in~\cite{zhou2013flexible}. The target GC-content is 0.5.  We compute the distance of the GC-content of the generated sequence with the target GC-content as a loss function for the GC-content. Then, finally, we sum up all of the normalized scores for different constraints and implement the Bayesian optimization in the latent space. 

By checking the training dequences, randomly generated RNA sequences with and without grammar and also after implementing the RGVAE and RNAGEN models, we filtered the sequences with the GC-content of $\pm 2\%$ of the target GC-content which has exactly the same RNA secondary structure of the target RNA secondary structure. No RNA sequences are found after the filtering in the training data, randomly generated RNA sequences with and without the grammar, and the generated sequences using RNAGEN model. But using the RGVAE model, eight RNA sequences are found after the filtering. Among these sequences, only one sequence could successfully satisfy the motifs and base positional constraints which are the nucleotides "UC" in the first two positions. This sequence is "UCACUUGGCUGUAGCU". Its GC-content is $50 \%$ and Its RNA secondary structure is "......(((....)))" which is exactly the target RNA secondary structure.

\subsection{RNA design with  alternative target structures}

In this section, we are going to optimize the generated RNA sequences for two target RNA secondary structures for the  105 nt \textbf{SAM} riboswitch with EMBL accession number AE016750.1/132874-132778 which is mentioned in~\cite{freyhult2007boltzmann}. This sequence is AACUUAUCAA GAGAAGUGGA GGGACUGGCC
CAAAGAAGCU UCGGCAACAU UGUAUCAUGU GCCAAUUCCA
GUAACCGAGA AGGUUAGAAG AUAAGGU. There are two different RNA secondary structures for this riboswitch as "..((((((.. .......... ((((.((((. .......((. ...)).(((( ......)))) )))).)))). .(((((.... .)))))...) ))))).."  and "..((((((.. ..(((((... (((.....)) )......))) ))(((.(((( ......)))) )))..(((.. .(((((.... .))))))))) )))))..". The energy for these RNA secondary structures are -28.10 and -26.7, respectively. To find an RNA sequence that has these two RNA secondary structures, first, we computed the base pair probability matrices for these two RNA secondary structures which contain elements 0 for unpaired nucleotides and 1 for paired nucleotides. Assume these matrices are $M_1$, and $M_2$, with the size $m \times m$. Then, we computed the base pair probability matrices for the generated sequences. Let us assume the matrix $N$ as one of these matrices with the size $n \times n$. After that, we computed the alignment between the base pair probability matrix of the generated sequence and the two base pair probability matrices of the target sequence by using a sliding window of the size of the smaller matrix over the diameter of the larger matrix. Then, we cropped the bigger matrix for the same size of the smaller matrix and computed the euclidean distance between upper triangles of these two matrices plus $|\frac{m^2-n^2 + m-n}{2}|$ which compensates the distance for the differences of the dimensions of the two matrices. Finally, we summed the score of the two target structures, minimized the score using Bayesian optimization, and compared the results with the training data and the sequences obtained using the RNAGEN model.

\begin{figure}
\begin{center}
\includegraphics[scale=0.5]{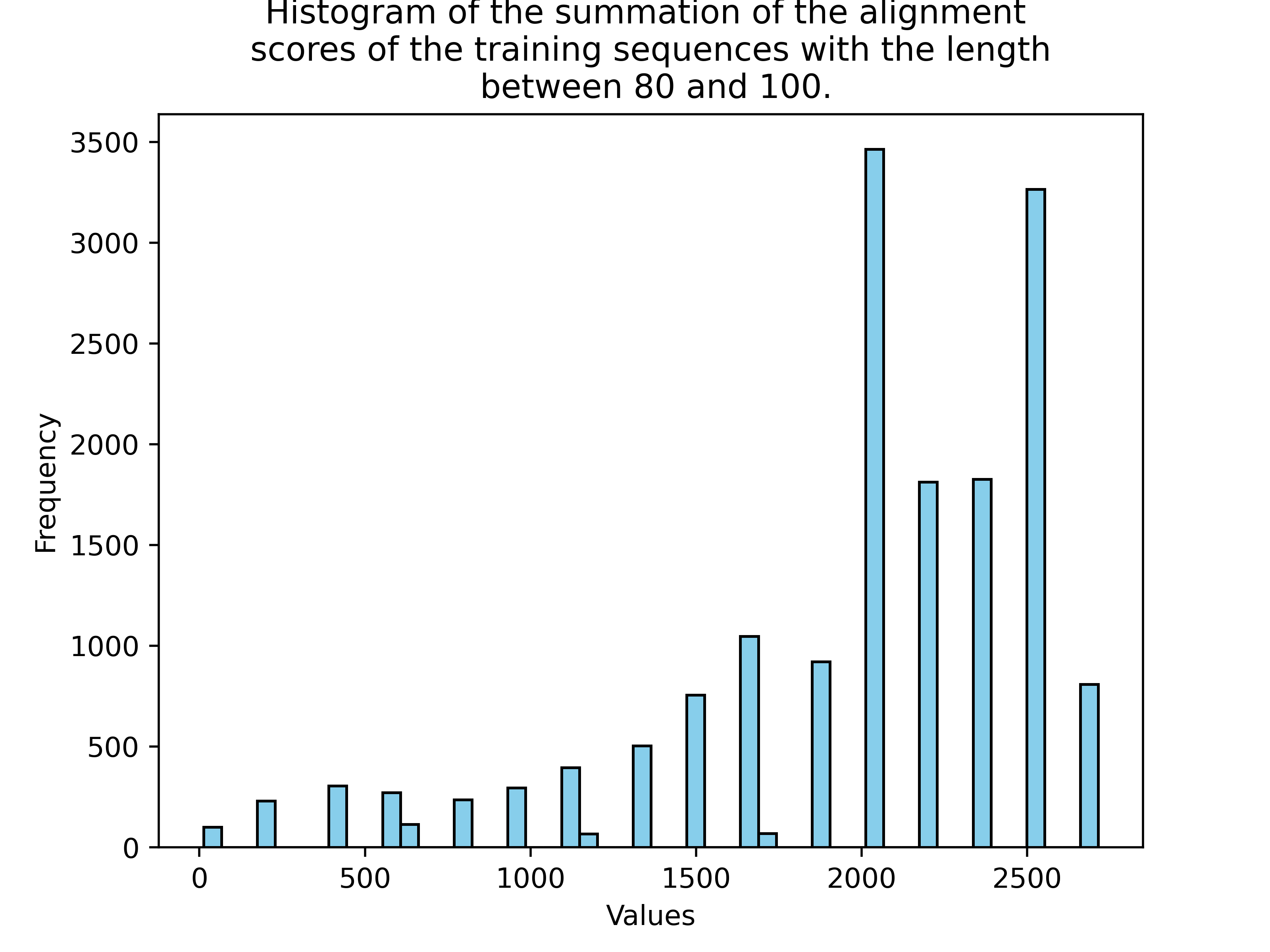}
\end{center}
\caption{Histogram of the summation of the alignment scores of the alignment scores of the sequences with the length between 80 and 100 for the training data. \label{aligntrain}}
\end{figure}

\begin{figure}
\begin{center}
\includegraphics[scale=0.5]{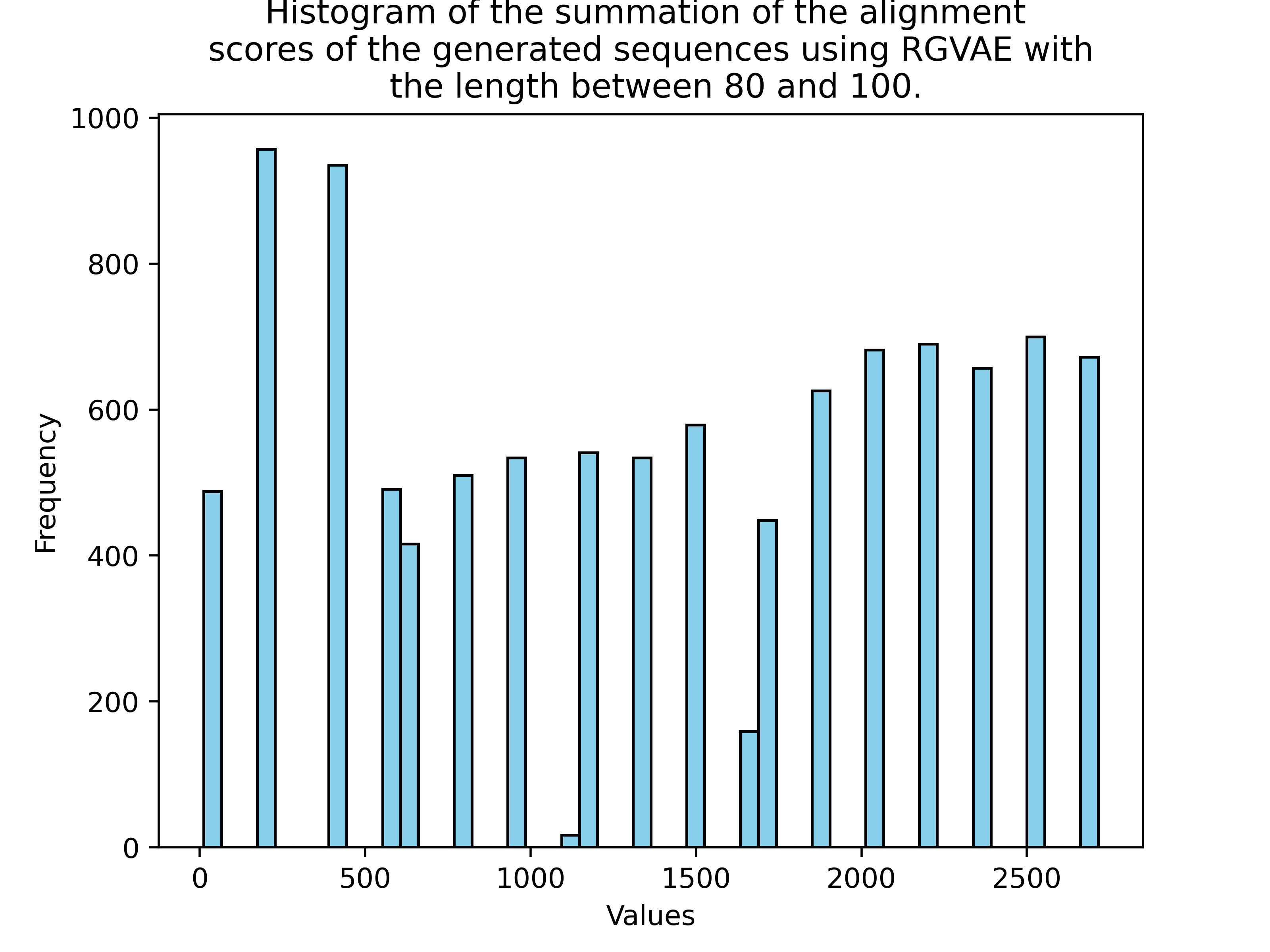}
\end{center}
\caption{Histogram of the summation of the alignment scores of the sequences with the length between 80 and 100 for the optimized proposed RGVAE model. \label{alignrgvae}}
\end{figure}

\begin{figure}
\begin{center}
\includegraphics[scale=0.5]{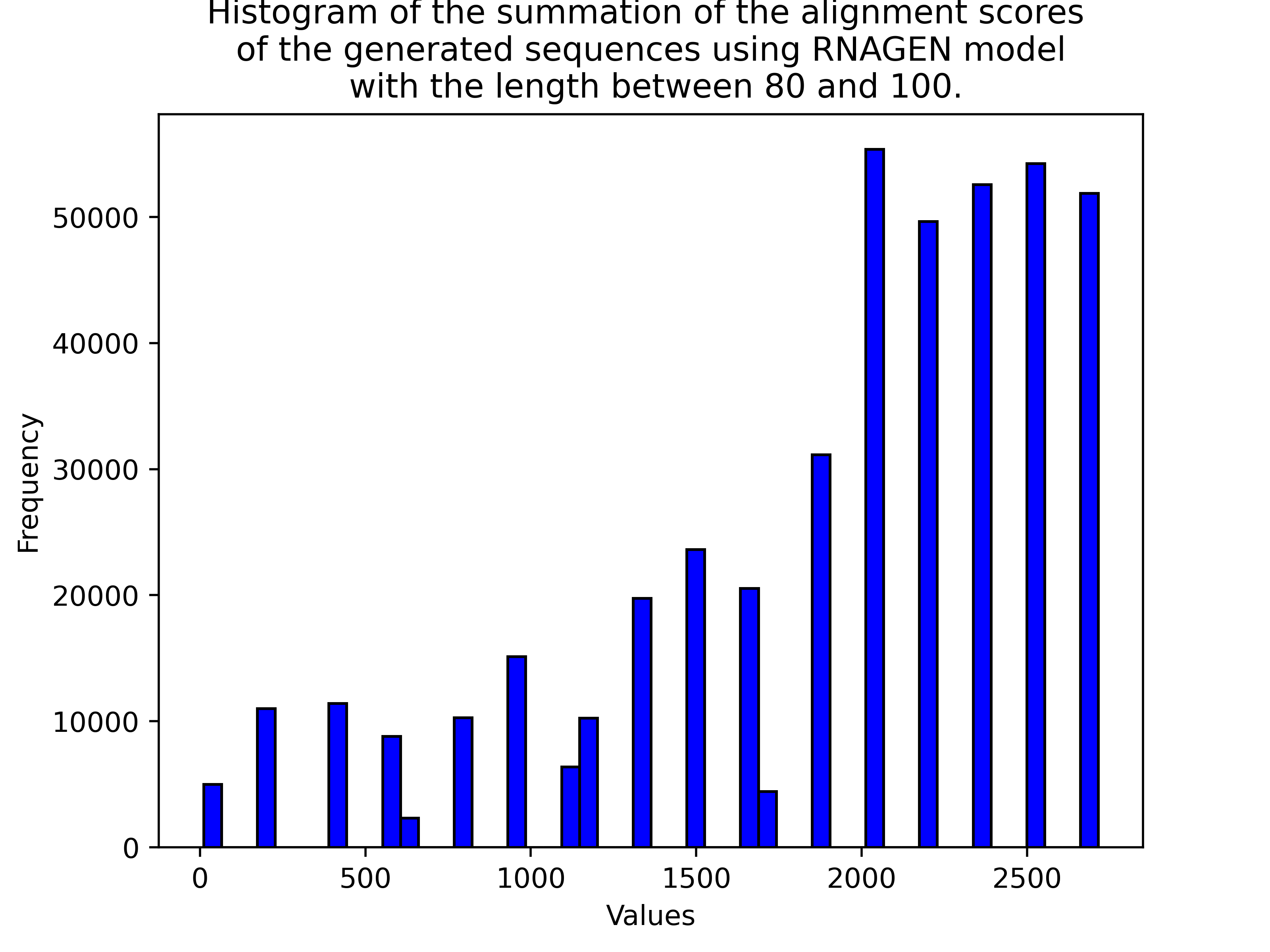}
\end{center}
\caption{Histogram of the summation of the alignment scores of the alignment scores of the sequences with the length between 80 and 100 for the proposed RNAGEN model. \label{alignrnagen}}
\end{figure}

Since the difference between the length of the generated or training sequences with the target sequence plays an important role in computing the distance, we selected the generated or training sequences with the length between 80 and 100. The histograms of the summation of the alignment scores for the training data, the optimized generated sequences using RGVAE, and the generated sequences using RNAGEN model with the length between 80 and 100 are shown in Figures \ref{aligntrain}, \ref{alignrgvae}, and \ref{alignrnagen}. As we see in these figures, the peak of the histogram for the optimized RGVAE model is lower than the training data and the generated sequences using RNAGEN model which shows that the generated sequences using RGVAE model with the length between 80 and 100 are more similar to the target structures of the riboswitch than the training data and the RNAGEN model.
Using the RGVAE, we found the best sequence with the minimum summation score as AAUUGCUGCC AAUAAUCUUC ACUUUGUUCA AAGUGAAUUU ACUGCGGAAA GGCUCCUUUC CGCAGAAAUC UAGGAAGAUA UAUUGGAGGC AGCAAU.
%AAUUGCUGCC AAUAAUCUUC GCUUUGUUCA AAGCGAAUUU ACUGCGGAAA GGCUCCUUUC CGCAGAAAUC UAGGAAGAUA UAUUGGAGGC AGCAAU.
%UGACAUUGCCUAACGCCUAGGCGUC. %AGACCACCAG UAUCCUACAG CUACGGAUCG #AAUGUUGAGC UGUUUCAUAG CCCCGACGCU #AUGAAAGGAC CAUUACAAUG GUUCAUGGUG #GUU. 
The sum of the distance of the alignment of the base pair probability matrix of this sequence with the base pair probability matrices of the two target RNA secondary structures is 11.47. We computed the suboptimal RNA secondary structures for the best generated sequence with the minimum summation distance. Then, we computed the hamming distance of this sequence with the target structures using sliding windows and added the length difference to it. We found the two alternative secondary structures of the generated sequence with the minimum hamming distance with the two target structures as "..(((((((( ........(( (((((....) ))))))(((( .((((((((( ......)))) )))))))))( (((....... ..)))).))) ))))).", and "..(((((((( ..((...((( (((((....) )))))))..) )((((((((( (....))))) )))))...(( (((....... ..)))))))) ))))).". 

%The energy of these two structures are -33.10, and -34.40 which are lower than the energy of the target 

%"....(.((((((.....))))))).", and ".(((...((((.(....))))))))". %The minimum of the corresponding value for training data is 68.28, for the randomly generated sequences is 85.80, and for the randomly generated sequences using the grammar is 83.68. The sequences found using the randomly generated sequences and the randomly generated sequences using the grammar are UGAUGUGGUU GAGGAGGGAC GUAGAUUGAA GGGCACCAAA UGAAAGCGGC CGGUCAGACC ACCAAGCUCA AACGAGGUUU UUCGUUUGGC GU and CCGGGACGCA CCAUAAAAUA UUGGGUUGGU CCCUGGUACG ACCGUAUCCA GCCCAUGAUU CUCGAAUCAU GGGGA, respectively. Hence, the results clearly show that RGVAE outperforms the randomly generated sequences and the training data to find the best sequence for two different RNA target secondary structures.

\section{Conclusions}
In this paper, we proposed RGVAE for generating and designing RNA sequences with specific constraints. We optimized the RNA sequences in the latent space of the RGVAE based on different combinations of constraints such as mandatory/forbidden motifs, minimum MFE, target GC content, specific target secondary structure, and base positional constraints. We trained our RGVAE model based on tRNA sequences, which have the well-known ``cloverleaf'' consensus secondary structure. As a result, the RNA sequences generated by the trained RGVAE are expected to possess similar structures that are thermodynamically stable. However, in case different types of secondary structures are desired, we may compose the training dataset for learning the RGVAE based on RNA families with consensus secondary structures whose characteristics resemble those of the target structure. For future work, we suggest to  optimize the latent space of the RGVAE to make the model more sampling-efficient for suggesting novel RNAs with multiple target properties~\cite{abeer2022multi}.

% \section{Equations}

% For footnotes in the main text of the article please number the footnotes to avoid duplicate symbols. \textit{e.g.}\ \texttt{\textbackslash footnote[num]\{your text\}}. The corresponding author $\ast$ counts as footnote 1, ESI as footnote 2, \textit{e.g.}\ if there is no ESI, please start at [num]=[2], if ESI is cited in the title please start at [num]=[3] \textit{etc.} Please also cite the ESI within the main body of the text using \dag.

\section*{Author Contributions}

Data Curation, Software, Methodology, Visualization, Writing – Original Draft (NZ);
Conceptualization, Funding Acquisition, Supervision, Writing – Review \& Editing (BJY); 
Methodology, Formal Analysis (NZ, BJY).

\section*{Conflicts of interest}

There are no conflicts to declare.

\section*{Acknowledgements}

For the numerical results of this research we have used the computing resources provided by Texas A\&M High Performance Research Computing (HPRC). In addition, we acknowledge stimulating discussions with Hossein Dehghani.

%Bibliography
\printbibliography
%\bibliographystyle{unsrt}  
%\bibliography{references}  

\end{document}